\documentclass[%
 reprint,
superscriptaddress,
 amsmath,amssymb,
 aps,
floatfix,
]{revtex4-2}

\usepackage{float}
\usepackage{graphicx}
\usepackage{dcolumn}
\usepackage{bm}
\usepackage{stmaryrd} 
\usepackage{amsmath}
\usepackage{hyperref}
\DeclareMathOperator{\Tr}{Tr}
\newcommand\sifigures{%
    \setcounter{figure}{0}
    \makeatletter 
       \renewcommand{\thefigure}{SI.\arabic{figure}}
    \makeatother}
\newcommand{\dropcap}[1]{#1}

\begin{document}

\preprint{APS/123-QED}

\title{Emergent circulation patterns from anonymized mobility data:\\ Clustering Italy in the time of Covid}
\author{Jules Morand}
\email{jules.morand@unitn.it}
\affiliation{Physics Department, University of Trento, via Sommarive, 14 I-38123 Trento, Italy, INFN-TIFPA, Trento Institute for Fundamental Physics and Applications, I-38123 Trento, Italy}
\author{Shoichi Yip}
\affiliation{Physics Department, University of Trento, via Sommarive, 14 I-38123 Trento, Italy, INFN-TIFPA, Trento Institute for Fundamental Physics and Applications, I-38123 Trento, Italy}
\author{Yannis Velegrakis}
\affiliation{Information and Computing Science, University of Trento, Italy and Utrecht University, Netherlands}
\author{Gianluca Lattanzi}
\affiliation{Physics Department, University of Trento, via Sommarive, 14 I-38123 Trento, Italy, INFN-TIFPA, Trento Institute for Fundamental Physics and Applications, I-38123 Trento, Italy}
\author{Raffaello Potestio}
\affiliation{Physics Department, University of Trento, via Sommarive, 14 I-38123 Trento, Italy, INFN-TIFPA, Trento Institute for Fundamental Physics and Applications, I-38123 Trento, Italy}
\author{Luca Tubiana}
\email{luca.tubiana@unitn.it}
\affiliation{Physics Department, University of Trento, via Sommarive, 14 I-38123 Trento, Italy, INFN-TIFPA, Trento Institute for Fundamental Physics and Applications, I-38123 Trento, Italy}


\date{\today}
\begin{abstract}
Using anonymized mobility data from Facebook users and publicly available information on the Italian population, we model the circulation of people in Italy before and during the early phase of the SARS-CoV-2 pandemic (COVID-19). We perform a spatial and temporal clustering of the movement network at the level of fluxes across provinces on a daily basis. The resulting partition in time successfully identifies the first two lockdowns without any prior information. Similarly, the spatial clustering returns $11$ to $23$ clusters depending on the period (``standard'' mobility \textit{vs.} lockdown) using the greedy modularity communities clustering method, and $16$ to $30$ clusters using the critical variable selection method. Fascinatingly, the spatial clusters obtained with both methods are strongly reminiscent of the 11 regions into which emperor Augustus had divided Italy according to Pliny the Elder. 
This work introduces and validates a data analysis pipeline that enables us: i) to assess the reliability of data obtained from a partial and potentially biased sample of the population in performing estimates of population mobility nationwide; ii) to identify areas of a Country with well-defined mobility patterns, and iii) to distinguish different patterns from one another, resolve them in time and find their optimal spatial extent. The proposed method is generic and can be applied to other countries, with different geographical scales, and also to similar networks (e.g. biological networks). The results can thus represent a relevant step forward in the development of methods and strategies for the containment of future epidemic phenomena.
\end{abstract}

\maketitle
\dropcap{I}n order to minimize the impact of epidemics such as the recent SARS-CoV-2 pandemic~\cite{Verity2020} on society, governments must take far-reaching decisions that considerably affect the lives of their citizens. Some common measures deployed during the pandemic were the adoption of personal protection devices such as face masks~\cite{Robinson2021, Talic2021}, contact tracing aimed at identifying and confining infectious subjects~\cite{Juneau2020, Ahmed2020, Liu2021, Colizza2021, Kostka2022, Ricci2021}, and the use of various forms of lockdown to dampen large-scale contagion~\cite{Alfano2020, Papadopoulos2020, Lavezzo2020, Nouvellet2021, Laio2022, Wallinga2004}. 

Italy was the first European country to impose a national lockdown and has seen the implementation of three nationwide lockdowns: between March and April 2020, in January 2021, and in April 2021. Detailed studies have been carried out on the initial propagation of the epidemic in Italy~\cite{Grasselli2020, Bertuzzo2020}, on the first confinement~\cite{Rinaldo2020}, and the relaxation of the latter~\cite{Merler2021}, discussing the necessity and the implementation of such restrictive measures. 

While lockdowns are certainly effective in curbing the rise of infections, their imposition severely affects the life and health of citizens~\cite{Urzeala2022, Gualano2020, Natilli2022}.
The extent of their deployment needs to be optimized both in space and time to minimize the number of people affected while guaranteeing the safety of the population. 
For this reason, after the first phase of the pandemic, the Italian government delegated part of the responsibility of restrictions to regional governments, which were forced to curb the movements of their citizens whenever the effective reproduction number $R_t$ (i.e. the average number of new infections caused by a single infected individual at time $t$) went above $1$~\cite{ISS_Rt,Gazzetta23Feb,Gazzetta2Marzo_1stConf}. 
Imposing regional lockdowns instead of national ones is a sensible strategy. However, it is not guaranteed that existing administrative regions correspond to the best subdivisions of a state to control the spread of epidemics. 

In general, a diffusion process in human society depends on the complex structure of the underlying network of interactions. At the individual scale, several studies make use of social experiments in recording the contacts of a group of people \textit{via} special devices; this was done e.g. in a summer camp for children in Italy~\cite{Cencetti2022} or with primary and high-school students in France~\cite{Fournet2014, Stehle2011, Barrat2013}. Such data can then be used to generate a time-dependent network of contacts that can be later used to simulate the diffusion of an epidemic and see how it develops at the scale of the single individual~\cite{Stehle2011a, Barrat2022}.
At larger scales, privacy concerns and pragmatic necessities can make it preferable to turn towards the usage of meta-population network models~\cite{Colizza2007, Colizza2008}. This can be done for example at a national~\cite{Unwin2020} or international level ~\cite{ColizzaBarrat2006,Le2022}, or at multiple levels through the usage of multiscale information on mobility~\cite{Balcan2009}. These and other studies can be informed by anonymized data such as airplane traffic~\cite{ColizzaBarrat2006,Le2022} or social network location data~\cite{Zhong2023}.

Here, by analyzing the mobility of the Italian population in the period between January 2020 and May 2022, we show how a data-driven meta-population approach can be used to identify the optimal spatial subdivision of a state to control an epidemics, as well as to verify a posteriori the effectiveness of lockdowns. To do so, we first estimate the mobility of the Italian population at the level of provinces (small administrative regions between municipalities and regions) thanks to Facebook (FB) data obtained through META's \textit{Data for Good} program~\cite{DataForGood}. To check the reliability of these data, we compute the population density vector  (i.e. the normalized vector of relative populations in the Italian provinces) obtained from META's data against the one derived from the independent data set of the Italian National Institute of Statistics (Istat~\cite{Istat}), containing the official projected census for January 1st, 2020~\cite{ISTATcensus}. The good agreement between these vectors shows that the data collected by META through the FB geolocalisation service provide a good estimate of the distribution of the Italian population. Then, we show how a clustering in time is able to correctly identify the first two national lockdowns, which were strictly enforced by the Italian state. Finally, we use the most characteristic mobility matrices for confined and non-confined phases to find the optimal spatial clustering of Italian provinces. To do so, we employ two different clustering methods that partition Italy into clusters of provinces matching with areas having cultural, social, and commercial affinities during 'ordinary' times, and into smaller clusters during confinement times. Applications of this approach to other Countries, scales, and other complex networks are discussed.
\ifdefined\nosecnumbers
\section*{Results}
\else
\section{Results}
\fi

Our approach to characterize the behavior of the Italian population is based on movement data between provinces. These are administrative entities in between regions and municipalities, usually containing between one and three hundred thousand people, with those corresponding to major cities such as  Rome, Naples, Milan, Turin, and Palermo having more than a million inhabitants~\cite{Istat}. 

As explained in detail in the Methods section, we consider 106 provinces and extrapolate the movement of their respective populations from FB users' data provided by META's data for good program~\cite{DataForGood}. The dataset we used provides the number of FB users in each province $i$, $n_i$, as well as the number of users moving between two provinces (or within a province), $n_{ij}(t)$, every 8 hours in the period between January 2020 and May 2022. 

\subsection{Transition matrices}

The data from META allow us to compute the 8-hours transition rate between two provinces $i$ and $j$,  defined as follows: 
\begin{equation}
    \Pi_{ij} (t) = \frac{n_{ij}(t)}{\sum_{j}n_{ij}(t)}.
\end{equation}

Note that the denominator ensures that, for every province $i$, $\sum_{j} \Pi_{ij} = 1$, thereby guaranteeing that $\Pi$ can be used as a stochastic matrix. 

\begin{figure}[ht]
    \centering
    \includegraphics[width=\linewidth]{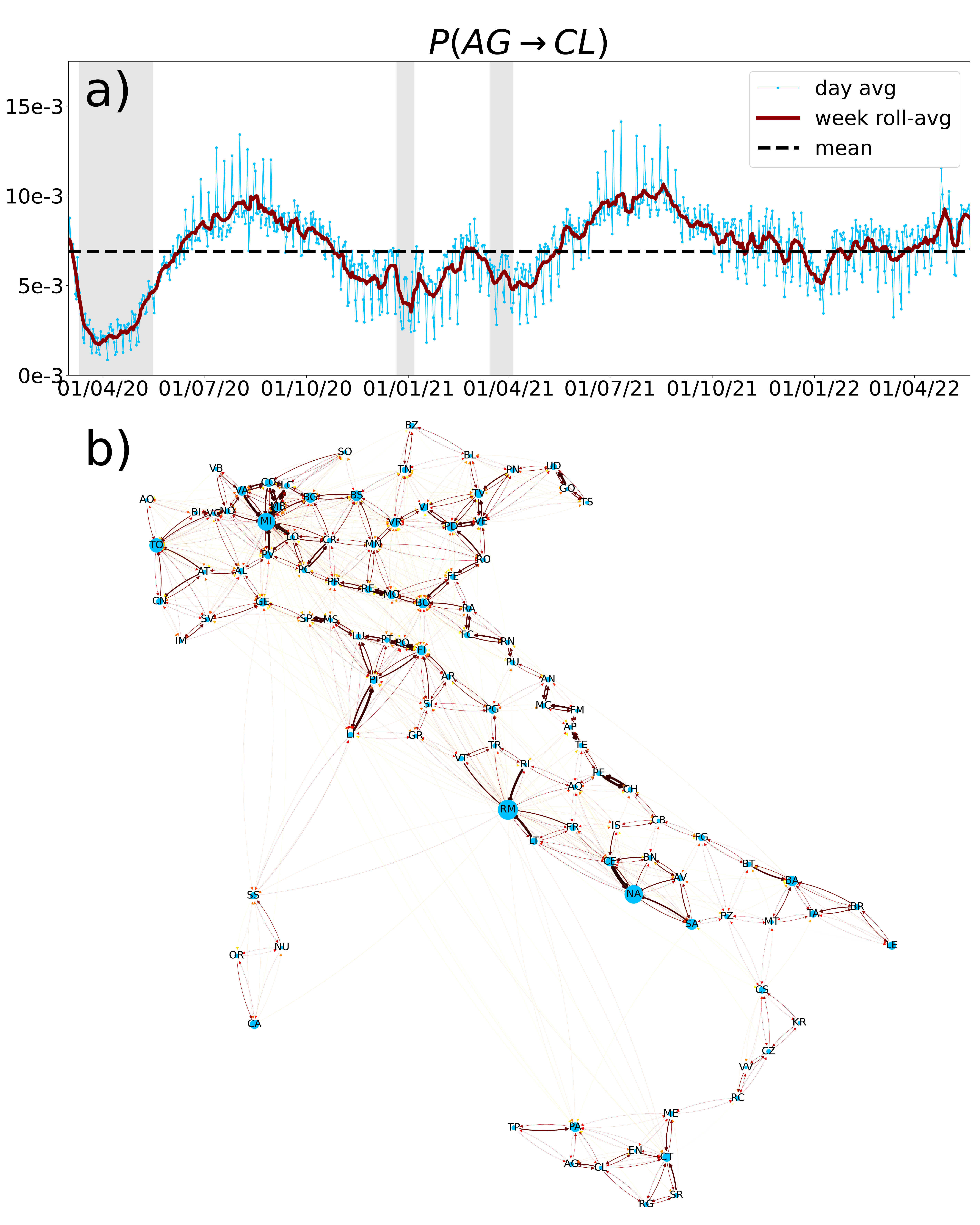}
    \caption{a) AG$\rightarrow$CL (Agrigento to Caltanissetta provinces) link vs time. Daily average probability (blue) and 7-day rolled-average probability (red), and overall probability averaged in time (black dashed line). b) Representation of the directed graph defined by the Matrix $\overline{\Pi}$ (Eq.\ref{eq:Pibar}). Arrows represent the mean probability links, $\overline{\Pi}_{ij}$, between Italian provinces $i$ and $j$, and are scaled in size and color according to the value of the link (from light yellow to dark red).
    The size of the nodes is proportional to the population (vector $\boldsymbol{\rho}^*$ of Fig.~\ref{fig:CompareNodesPop_FB_Usage}). Self-links $\Pi_{ii}$ are not shown as they have a much bigger value  ($\sim \times 10^2$) than the non-diagonal links, as most people do not move out of their province.}
    \label{fig:Link_VS_Time-GraphItalie}
\end{figure}

To get an idea of what the data look like, the time evolution of one link $\Pi_{ij}$, reporting the mobility from the province of Agrigento ($i=AG$) to that of Caltanissetta ($j=CL$), is plotted in Fig.~\ref{fig:Link_VS_Time-GraphItalie}a). Daily averaged values are reported in blue, weekly averaged ones in red, and the corresponding entry in the mean matrix of Eq.~\ref{eq:Pibar} (see materials and methods) in a black dashed line. The lockdown periods are indicated by grey-shaded vertical bars. Seasonal effects are clearly visible from the comparison of the daily data and the corresponding weekly averaged ones.

To remove seasonal fluctuations in $\Pi$ (day vs night, weekdays vs weekends) we redefine $\Pi$ as the daily transition rate between provinces averaged over the three days before and three days after, see Materials and Methods section. Finally, it is convenient to consider the mean transition matrix over the whole period, $\overline{\Pi}$. The directed graph associated with $\overline{\Pi}$ is displayed in Fig.~\ref{fig:Link_VS_Time-GraphItalie}b). 

\begin{figure*}
    \centering
    \includegraphics[width=0.95\linewidth]{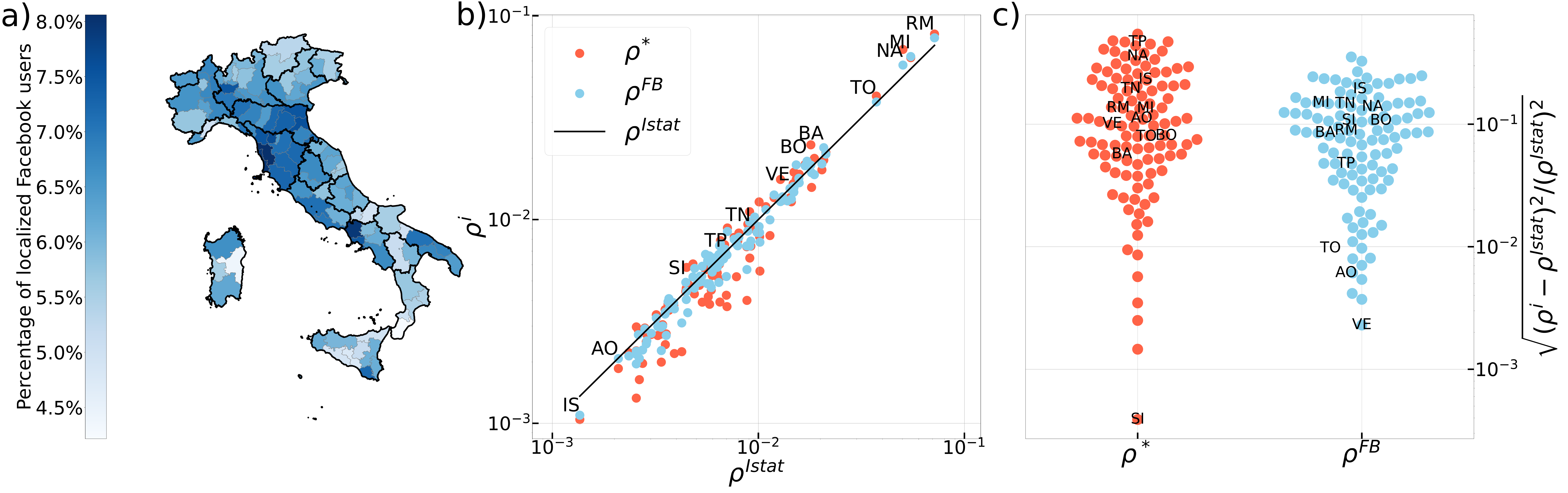}
    \caption{a) Fraction of FB users that have shared their location over the official province population obtained from the Istat 2020 census, $\overline{n_i}/n_i^{\text{Istat}}$, for each province $i$. b) Comparison of the different population density vectors from FB and Istat data: $\boldsymbol{\rho}^{\text{FB}}$ and $\boldsymbol{\rho}^{*}$ are plotted against $\boldsymbol{\rho}^{\text{Istat}}$. c) Standard deviation of the vectors $\boldsymbol{\rho}^{\text{FB}}$ and $\boldsymbol{\rho}^*$ from the $\boldsymbol{\rho}^{\text{Istat}}$  vector.  }
    \label{fig:CompareNodesPop_FB_Usage}
\end{figure*}

\subsection{Homogeneity and representativeness of FB data}\label{sec:PopulationDensityVectors}

We assume that the FB users in the database are homogeneously distributed across provinces, and move in a manner that is on average similar to that of the rest of the population. To validate these assumptions we proceed as follows.

First, we monitor the fraction of FB users over the total population of the province according to Istat; this ratio is defined as  $\overline{n_i}/n_i^{\text{Istat}}$, where $\overline{n_i}=\left<n^h_i\right>$ is the number of FB users in province $i$ averaged over the whole time series. The results, reported in Fig.~\ref{fig:CompareNodesPop_FB_Usage}a), show that in all provinces this fraction remains between $3\%$ and $7\%$, and that FB users are roughly homogeneously distributed across the country (Fig.S2 of supporting information displays the vectors with all provinces two-letter codes).

A more quantitative validation of both assumptions can be obtained by considering the \emph{population density vectors} obtained both from the official census of Istat in 2020 and from FB users' data. These are defined as follows:
\begin{equation}\label{eq:popdenvec}
    \boldsymbol{\rho} = \left(\frac{n_{1}}{n_{tot}},\ldots, \frac{n_{N}}{n_{tot}}\right)^T
\end{equation}
where $n_1, \ldots n_{N}$ are the populations of the $N$ provinces, and $n_{tot} = \sum_{i=1}^{N} n_{i}$ is the total population. 
The populations $n_{i}$ can be obtained from either:
\begin{itemize}
    \item Istat data, $\boldsymbol{\rho}^{\text{Istat}}$,
    \item the FB population dataset, $\boldsymbol{\rho}^{\text{FB}}$,
\end{itemize}

The above normalization, Eq.~\ref{eq:popdenvec}, sets $|\boldsymbol{\rho}|=1$ and allows us to compare the different vectors. In addition, our approach makes it possible to compare another population density vector, $\boldsymbol{\rho}^*$, obtained from the mean transition matrix $\overline{\Pi}$ extracted from the FB movement dataset. 

In the graph described by $\overline{\Pi}$ there is a non-zero probability to reach any node from any other one in a finite number of steps, that is, the graph is strongly connected and aperiodic, and random walks over it are ergodic. The Perron-Frobenius theorem then ensures that $\overline{\Pi}$ has a non-degenerate highest eigenvalue. With our normalisation of $\overline{\Pi}$ this is $\lambda^*=1$, and its associated left eigenvalue $\mathbf{\rho}^*$ is the only stationary state of the system, satisfying: 
$$\rho^*_i \overline{\Pi}_{ij}= \overline{\Pi}_{ji} \rho^*_j.$$
Therefore, any non-trivial distribution  vector over the nodes of our network will converge to $\boldsymbol{\rho}^*$ after a sufficiently long time (see supporting information: section 3)

If the movements described by $\overline{\Pi}$ are consistent with the Istat population data, the stationary density vector $\boldsymbol{\rho}^*$ must be in good agreement with the Istat density vector $\boldsymbol{\rho}^{\text{Istat}}$. This is indeed the case, as shown in Fig.~\ref{fig:CompareNodesPop_FB_Usage}b) and c).

Fig.~\ref{fig:CompareNodesPop_FB_Usage}, panel b) displays the population density vectors $\boldsymbol{\rho}^{\text{FB}}$ and $\boldsymbol{\rho}^{*}$,  on a log-log scale against $\boldsymbol{\rho}^{\text{Istat}}$. The provinces are sorted from least to most populated according to Istat data. We see a good agreement within the FB data themselves, which is also a benchmark of our extraction and preparation of the data.

Moreover, the standard deviations of $\boldsymbol{\rho}^{\text{FB}}$ and $\boldsymbol{\rho}^{*}$ from the Istat vector (panel c of Fig.~\ref{fig:CompareNodesPop_FB_Usage}) are in very good quantitative agreement with the Istat data. However, we notice that the most populated provinces, Rome, Milan, Naples, Turin, (RM, MI, NA, TO) are slightly overestimated and that the less populated provinces are slightly underestimated especially by the $\rho^*$ vector. This can be explained by the fact that all links with less than 10 people are ignored for privacy reasons.

\subsection{Transition matrices time series}\label{sec:timeseries}

\begin{figure}[t]
    \centering
    \includegraphics[width=\linewidth]{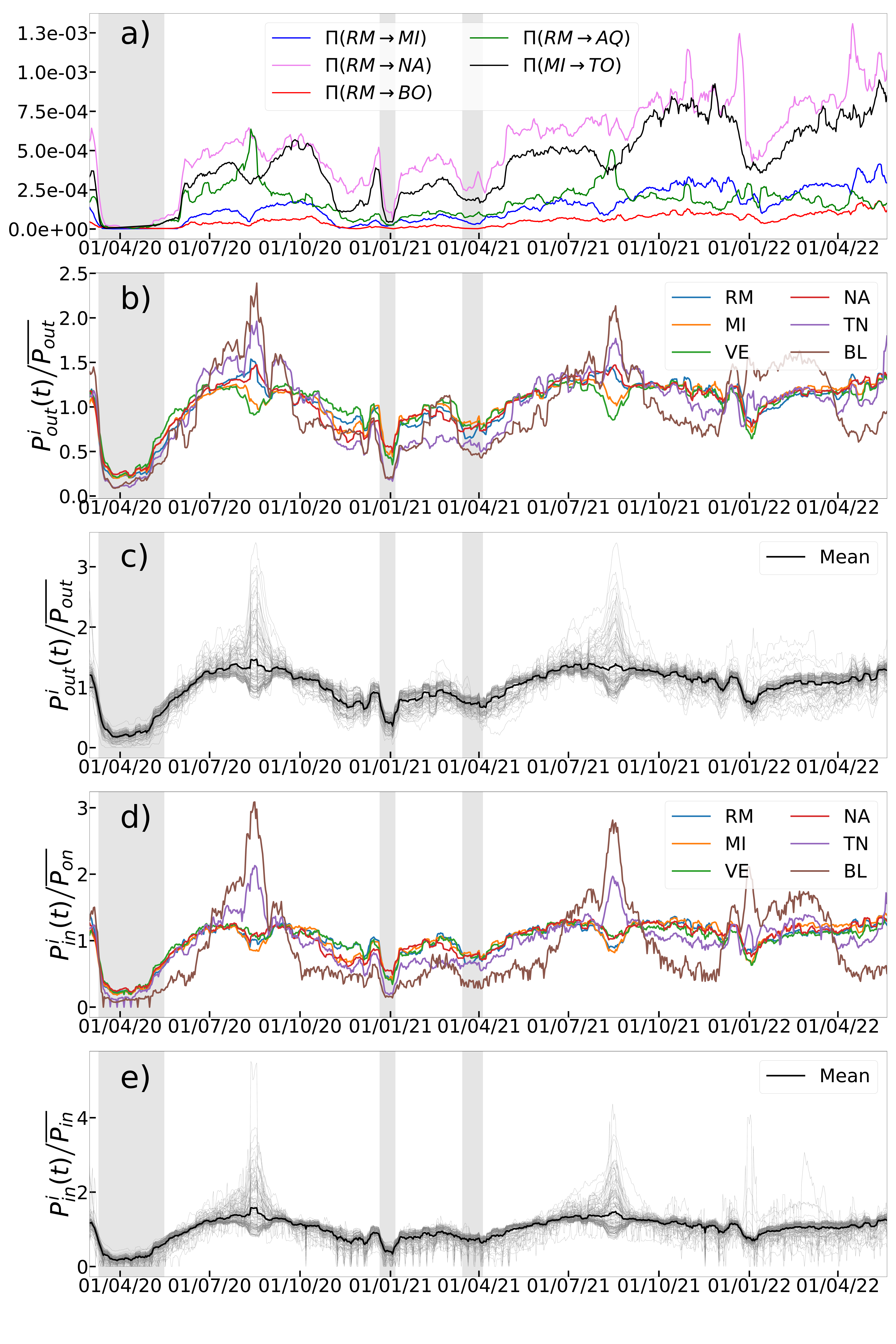}
    \caption{
    a) Examples of some representative transition probability links. 
    b) and c) Probability of going out of the province versus time b) for $i=$BL, MI, NA, RM, TN, and VE and  c) for the mean over the province in black and the whole distribution of grey. 
    d) and e) Probability of going in the province versus time d) for $i=$BL, MI, NA, RM, TN, and VE  and e) for the mean over the province in black and the whole distribution of grey. All probability distributions have been plotted and re-scaled by their temporal average to obtain a collapse of the curves. Gray shaded areas represent national lockdown periods.}
    \label{fig:LinkCompare}
\end{figure}

Having validated the FB data, we proceed to extract the information contained in the time series of weekly-averaged daily transition matrices. First of all, we notice that diagonal elements $\Pi_{ii}\geq 0.9$, meaning that most movements happen within provinces. Second, and most notably, we find that while the time series of the probability to move between different provinces can vary by an order of magnitude, as shown in Fig.~\ref{fig:LinkCompare}a), the movement pattern of single provinces can be brought to collapse on two master curves with an appropriate rescaling, see Fig.~\ref{fig:LinkCompare}b)-e). Specifically, this can be done by considering the normalized probability to move out of a province,
\begin{equation}
   \left. P_{out}^i(t)\right/\overline{P_{out}}=\left. (1-\Pi_{ii}(t) \right/ (1-\overline{\Pi}_{ii}),
\end{equation}
shown in Fig.~\ref{fig:LinkCompare}b), c) and the normalized probability to move into a province,
\begin{equation}
   \left. P_{in}^i(t) \right/\overline{P_{in}}=\left.\sum_{j\neq i}\Pi_{ji}(t)  \right/ \sum_{j\neq i}\overline{\Pi}_{ji}, 
\end{equation}
reported in Fig.~\ref{fig:LinkCompare}d), e). As can be seen from panels c), e) all provinces display a similar behavior in these two quantities, and the first two lockdowns become apparent as periods of low mobility.
The Z-score, i.e. the time average of the fluctuation of $P_{in}^i(t)$ and $P_{out}^i(t)$ with respect to the mean over provinces, is defined and displayed in supporting information (Fig.S6).

Interestingly, we also note that some provinces show a large deviation in both quantities in correspondence of summer and winter months. To rationalize this, we look at the provinces showing peaks of mobility in those periods, and found them to correspond with those having a high touristic vocation, as for example Belluno (BL) and Trento (TN), near the Dolomites, and Sicilian provinces, see Fig.~\ref{fig:LinkCompare}b), d). While at first the fact that Rome and Venice (VE) do not show these peaks might be unexpected, we recall that our data only follow the movement of Italian citizens, and that during Covid there was a strong push to take holidays outside of cities.

\subsection{Temporal Clustering}\label{sec:ResultsTempClus}
\begin{figure}[ht]
    \centering
    \includegraphics[width=\linewidth]{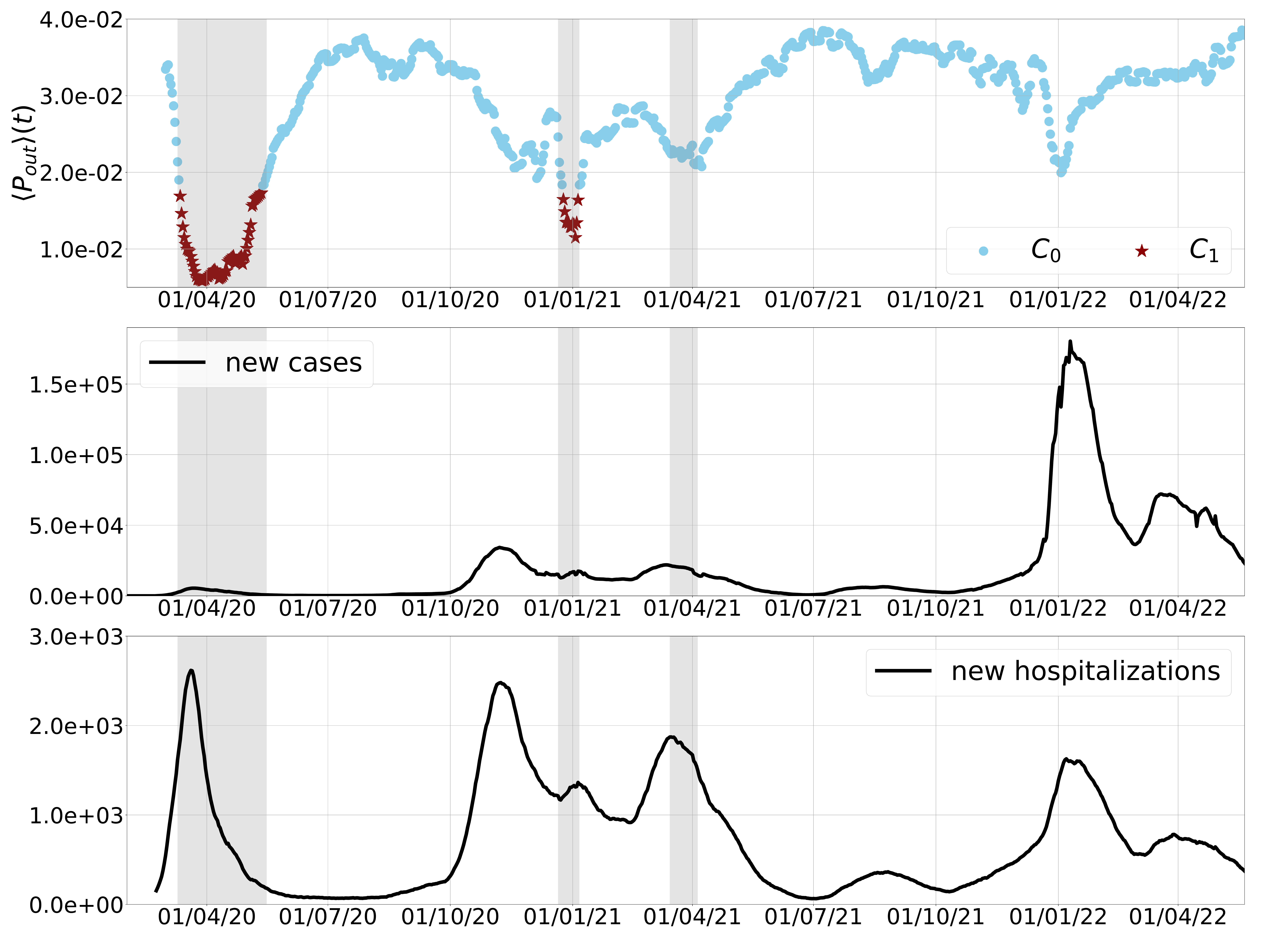}
    \caption{Top panel: Mean mobility $\langle P^{out}\rangle(t)$ versus time. The light blue dots and dark red stars illustrate the two temporal clusters of transition matrices series. Gray-shaded areas represent national confinement periods. 
    Center panel: Number of new cases of COVID-19 per day in the whole of Italy versus time. Bottom panel: Number of new hospitalized cases due to COVID-19 in Italy per day.}
    \label{fig:TemporalClustering}
\end{figure}
In order to reach our goal, that is to use movement data to identify spatial communities, we first need to ensure that the information contained in the transition matrices $\Pi(t)$ is sufficient to identify the national lockdown periods.

To do this, we cluster the daily movement matrices into two groups based on the distance induced by the matrix-matrix scalar product, as described in the Materials and Methods section. 
The results are reported in the top panel of Fig.~\ref{fig:TemporalClustering}, where each matrix is represented by the average probability for people to move out of their province at time $t$: 
\begin{equation}
\left<P_{out}\right>(t)=\frac{1}{N} \sum_{i=1}^N 1-\Pi_{ii}(t)= 1-\frac{1}{N}\Tr(\Pi(t)).
\end{equation} 

The two temporal clusters $C_0$ and $C_1$ are represented by light blue dots and dark red stars, respectively, and the latter clearly identifies the first two national lockdowns periods, delimited by the vertical shaded areas. Although the third lockdown period is not identified by the clustering, we argue that this is because it has not been strictly imposed, nor was it effectively respected, as can also be seen from the mobility plots of Fig.~\ref{fig:TemporalClustering}, and Fig.~\ref{fig:LinkCompare}b)-e).

%

While in order to fully understand the behaviour of the new infections and hospitalization curves, one would need to take into account a number of factors, such as for example population density and temperature variations~\cite{Smith2021}, the repercussions of the confinements on the evolution of the epidemics are clearly visible in Fig.~\ref{fig:TemporalClustering}: the lockdowns are all followed by a decrease in the number of new cases and new hospitalizations, as expected~\cite{Nouvellet2021,Lavezzo2020}. Furthermore, we can notice that the curves for $\langle P_{out}(t)\rangle$, new cases, and new hospitalizations, are in general anticorrelated, with mobility decreasing in correspondence to increases in the other two curves, which then reach a peak and decrease. The reduction in mobility outside of national lockdowns is arguably due to individual decisions and even more to local movement restrictions applied by regions; the fact that a lower mobility leads to a decrease in the number of infections is a standard prediction of epidemic models.

FB data thus entail mobility features that are in agreement with the history of the Italian government's decisions and their repercussions on the population's behaviour, validating their usage in modeling epidemics and social phenomena more in general.  

\subsection{Optimal Spatial Clustering}\label{sec:OptimalSpatialClustering}
\begin{figure*}[ht]
    \centering
    \includegraphics[width=\linewidth,trim={0 0.5cm 0 0.2cm},clip]{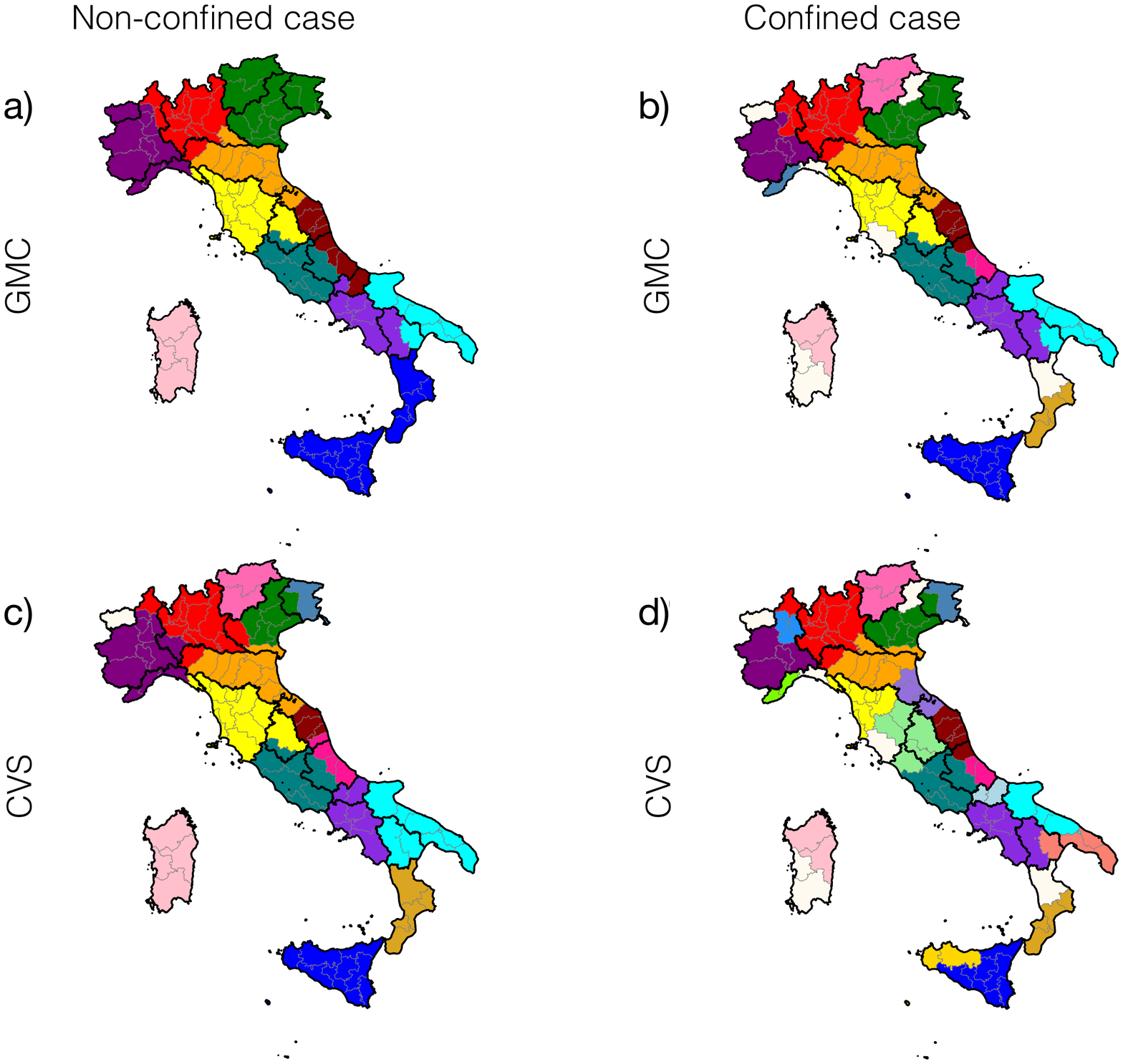}
    \caption{Spatial community clustering obtained with Greedy Modularity, a), b), and Critical Variable Selection, c), d). Panels a), c) report the communities identfied by the two methods during the non-confined periods. Panels b), d), during confinement. Grey lines represent the borders of the provinces while bold black lines delimit administrative regions.}
    \label{fig:Maps}
\end{figure*}

We can now perform a spatial clustering of the most representative matrices of the two temporal clusters obtained for the confined and unconfined situations.\\
To this aim, we define for each temporal cluster ($C_k, \, k=0,1$):
\begin{itemize}
\item the \emph{mean transition matrices} $\overline{\Pi}^{C_k}$ , 
\item the \emph{most representative transition matrices} $\widetilde{\Pi}^{C_k}$,
\item the \emph{most representative current matrices} $J^{C_k}\!\!=\widetilde{\Pi}^{C_k}\rho^{\text{Istat}}$. 
\end{itemize}
We then use two different methods to perform the clusterization itself:
\begin{itemize}
    \item the Greedy Modularity Communities (GMC) method, which uses $J^{C_k}$, i.e. the flux of people moving, and finds the number of clusters that maximizes \emph{modularity}, a concept from graph theory. This algorithm optimizes the clustering such that the inner links of clusters are stronger than the outer ones.
    \item the critical variable selection (CVS) method, which makes use of $\widetilde{\Pi}^{C_k}$, i.e. the probability of a single person to move, and finds the number of clusters that maximizes the \emph{relevance}, a quantity introduced in information theory. This method searches for the clustering that minimizes information loss with respect to a full description of the dataset~\cite{Marsili2013}.
\end{itemize}
The details of both strategies are reported in the Materials and Methods section and graph representation of the most representative matrix in each case can be found Fig.S9 and Fig.S10 of supporting information. We observe here that, although in principle geographically distant provinces could be grouped together (e.g. in the case of highly-connected cities such as Rome, Naples, Milan, and Turin), the clusters found by both methods are composed of physically proximal provinces, which can be reached one from the another without having to cross other clusters. This is a non-trivial result, as neither method relies on the notion of geographical distance. 

\subsubsection*{Non confined}

Fig.~\ref{fig:Maps}a),c) represent the clustering of the most representative matrix of the unconfined temporal cluster ($C_0$ in blue in the top panel of Fig.~\ref{fig:TemporalClustering}), corresponding to an `ordinary' Italian mobility situation; the top map is obtained employing the greedy modularity method, while the bottom one makes use of the CVS approach.

The two methods return slightly different partitions: for the greedy modularity (top), the Italian provinces are grouped in $11$ clusters corresponding to well-defined geographical areas, while $16$ groups are found using the CVS scheme. 
Moreover, apart from a few border cases, the clusterings almost perfectly reproduce well-known cultural and commercial `blocks' within the Country. 
For example, the green cluster corresponds to the \textit{Triveneto} area (that is Veneto, Friuli-Venezia Giulia, and Trentino-Alto Adige), while Sardinian provinces are fully grouped in their own cluster. 
The time series of outward and inward probabilities for each province are also displayed in the supporting information (Fig.S7 and Fig.S8) with a highlight for each optimal spatial cluster obtained with the greedy modularity method.

It is interesting to observe that the $11$ clusters found through modularity resemble quite strongly those reported by Pliny the Elder~\cite{Pliny,Cifani2010}, according to whom Emperor Augustus divided Italy into $11$ regions around 7 BC. The comparison is reported in the supporting information (Fig.S3-S4).

\subsubsection*{Confined}

Things change dramatically when the matrix representing the confined case ($C_1$: cluster 1, in red in the top panel of Fig.~\ref{fig:TemporalClustering}) is considered. Fig.~\ref{fig:Maps}b), d)  display the corresponding clustering, in the top panel using the greedy modularity method and in the bottom one using CVS. In this case, the optimal clustering produces $23$ spatial clusters with the former approach and $30$ with the latter. Both of them predict more clusters, as expected when mobility is reduced.  By analyzing the most representative matrices as directed graphs, one can also see that the one for the confined case presents fewer links than the one for non-confined mobility, and that some provinces become singletons in the optimal spatial clustering, see supporting information (Fig.S9 and Fig.S10).

Also in this case both clustering methods provide comparable results: most of the north of Italy is partitioned similarly with both methods; the singletons (off-white) are essentially the same; also Trento (TN) and Bolzano (BZ), Sassari (SS) and Nuoro (NU), as well as Pescara (PE) and Chieti (CH) are clustered together by pairs with both methods.

\ifdefined\nosecnumbers
\section*{Conclusions}
\else
\section{Conclusions}
\fi
Picking the period 2020-2022 in Italy as a test-case, we introduced a method to assess the main patterns and the most representative movements in a state by using anonymized data from social networks, and showed how this information can be used to identify temporal patterns and spatial communities.
The temporal links of the network were inferred from the movement dataset of Facebook users provided by META through its \textit{Data for good} program, combined with publicly available databases from Istat and ISS. 

We showed how movement data from social networks can be validated by considering the associated average transition matrix between provinces as the generator of a Markov jump process, and comparing the corresponding stationary density vector with the population density vector obtained from the official census. 


By analyzing the transition matrices time series, we then showed how the normalized probabilities of people moving out of or into provinces collapse on the same average curves, irrespective of the resident populations. Deviations from these two curves can be used to identify 
particular fluxes due e.g. to tourism. 

By considering the distance between transition matrices, we were then able to perform a temporal clustering to distinguish the lockdown periods from the rest. This successfully identifies the first two national lockdowns, which were strictly enforced by the Italian Government. Finally, we picked the most representative transition matrices from the confined (lockdown) and non-confined periods, and used two different methods to identify spatial communities: greedy modularity communities and critical variable selection. The first one finds the communities whose populations move more within a cluster than between clusters; the latter defines clusters to reduce the distance traveled by individuals based on an information-theoretical approach. 
Both methods return an optimal scale at which actions on circulation in a country could be enforced, and their results are consistent with one another.
%
%
%

As our methodology is completely general, these strategies can be applied to other countries or other scales, as well as different problems relying on a similar kind of data, such as optimizing a transportation network in a city \cite{Bontorin2023}, or the analysis of the interaction network between residues to identify coherent or persistent structure in protein dynamics~\cite{tiberti2014pyinteraph}.


Finally, we point out that our strategy could also be used to compare the current socio-economical communities in a country to historical data. This can be particularly interesting in regions with a long record of historical documents and whose borders changed significantly over time, of which Italy is a prime example. In this respect, it is noteworthy to observe that the non-confined communities found by modularity  bear a strong resemblance with those reported by Pliny the Elder in its Naturalis Historia~\cite{Pliny}, 
suggesting that the current social, economical, and mobility pattern of Italian communities still echoes its roots dating back by almost two millennia.



\section{Materials and methods}
\let\section\subsection
\let\subsection\subsubsection
\section{Datasets}\label{sec:data}
\subsection*{Facebook movement data}
The Facebook (FB) movement data were taken from META's \textit{Data for good} program. The database records the number of people going from province $i$ to province $j$, updated every $8$ hour, for Italian users who allowed FB to share such information with the app on their device; the time frame covered goes from March 1st, 2020 to May $22$nd, 2022 ($811$ days). The database has been completely anonymized by META~\cite{MetaPrivacy}. In particular, all links between two provinces containing less than $10$ people are ignored.\\

The FB movement data are available both on a grid with cells of roughly $600 \times 600$ meters at the equator, which is the minimum tile size allowed for privacy protections (Bing tile level 16~\cite{Bing}), and at the scale of Italian provinces, administrative entities in between municipalities and regions.
In this study we concentrate on the province level: the list of $106$ provinces used was the official one in 2016 except for the provinces of Sud Sardinia (SU) and Cagliari (CA) which were merged into one node (CA), in order to get inter-compatibility of administrative regions between datasets from FB, Istat, and ISS. A map (Fig.S1) and a table of these provinces can be found in the section 1. of supporting information. In section 2. of section of supporting information describes in detail the workflow of the data preparation.

In this database the FB data reports for each $8$ hour period (labeled by $h$):
\begin{itemize}
    \item The number of FB users moving from province $i$ to province $j$ at time $h$, $n^h_{ij}$ (called $n_{crisis}$ in the original dataset).
    \item The total number of FB users in province $i$ at time $h$, $n^h_i$. 
\end{itemize}

\subsection*{Istat and ISS data}
The FB data cover only a fraction of the Italian population (namely those individuals who employ the FB app on mobile devices and enabled location sharing) and does not provide direct information on the population of each province, the amount of COVID cases registered there, or the duration of confinement periods. The population of each province $i$, $n^{\text{Istat}}_i$, was obtained from Istat~\cite{Istat}, the Italian National Institute of Statistics. We used the most recent database available before the pandemic, released on January 1st, 2020. For simplicity, we assumed that the population remained constant during the period of study: this is an acceptable approximation, given that the global growth rate of the Italian population for that period is roughly $-0.4\%$~\cite{MacroTrends} and this fluctuation is negligible for our analysis.

The amount of new COVID cases between February 1st, 2020, and October 7, 2022, reported in the bottom panel of Fig.~\ref{fig:TemporalClustering}, was obtained from ISS~\cite{ISS}. Data were accumulated as a rolling average over one week.

The dates of the national confinements implemented by the Italian government are the following~\cite{Gazzetta2Marzo_1stConf,Gazzetta2Dicembre_2ndconf,Gazzetta13marzo2021_3rdconf}:
%
\begin{enumerate}
    \item from 10/03/2020 to 16/05/2020;
    \item from 21/12/2020 to 06/01/2021;
    \item from 15/03/2021 to 05/04/2021.
\end{enumerate}

The three periods are indicated by the grey-shaded areas in Figs.~\ref{fig:TemporalClustering} and ~\ref{fig:LinkCompare}. The confinement and de-confinement were progressive processes e.g. at first not all provinces were confined: only two days after the initial, local lockdown the measure was applied to the whole Country. Hence, we chose the temporal boundary of the lockdowns such that the periods correspond to the situation where the whole Country was confined, particularly periods in which any movement between provinces was prohibited.

At smaller scales, national confinements were characterized by rigid restrictions on mobility: in particular, the Italian government provided sanctions of up to three months in prison for those who violated the lockdown, and all non-essential facilities and shops were closed, gyms, swimming pools, spas, wellness centers, museums, cultural centers, ski resorts, cinemas, theatres, pubs, dance schools, game rooms, betting rooms, and bingo halls, discos and similar places in the entire country were suspended. All organized events were also suspended, as well as events in public or private places, counting those of cultural, recreational, sporting, civil, and religious ceremonies, including funeral ceremonies~\cite{Gazzetta2Marzo_1stConf}.

\section{Stochastic transition matrices}

Using the data described in Sec.~\ref{sec:data}, we built the transition matrices between provinces. As described below, these are averaged daily and over the whole period.


\subsection*{Mean transition matrix over the whole period}

FB data allowed us to define a mean transition matrix $\overline{\Pi}$ between nodes as follows:
\begin{equation}
\label{eq:Pibar}
\overline{\Pi}_{ij}=\frac{\sum_h n^h_{ij}}{\sum_j\sum_h n^h_{ij}}    
\end{equation}
where $\sum_h$ is the sum over all 8-hour-slots during the whole data period. The denominator in Eq.\ref{eq:Pibar} normalizes the matrix such that the elements in each row sum to one: $\sum^N_j \overline{\Pi}_{ij}=1, \forall i$, thus ensuring that $\overline{\Pi}$ is a stochastic matrix.
 

\subsection*{Daily transition matrix}\label{sec:dailytrans}
FB data were used to generate a daily transition matrix representing the link between provinces for each day, indexed by $t$. The time evolution of the mobility network was monitored by constructing a time series of transition matrices as follows:
\begin{equation}
\label{eq:DefintionPijt}
\Pi_{ij}(t)=\frac{\sum_{h\in [t - \epsilon,t+\delta]} n^h_{ij}}{\sum_j\sum_{h\in [t - \epsilon,t+\delta]} n^h_{ij}}   
\end{equation}
where $\sum_{h\in  [t - \epsilon,t+\delta]}$ is the sum over all 8-hours-slots in $[t - \epsilon,t+\delta]$. 

Using Eq.~\ref{eq:DefintionPijt} we constructed two different daily time series, one averaged every 24 hours, $\epsilon =0$, and $\delta = 24$h, and one based on a weekly rolling average, $\epsilon = 72$h days, $\delta = 96$h (in between 3 days before and 3 days after day $t$). The weekly averaged one correspond to the average of data provided by ISS.



\section{Temporal Clustering method}\label{sec:TempClusMethods}
%
To perform the temporal clustering of the transition matrices $\Pi(t)$, we used the standard Frobenius matrix norm:
\begin{equation}\label{eq:Fnorme}
\Vert \Pi(t) \Vert = \Tr(\Pi(t).\Pi(t)^T)
=\sqrt{\sum_{i=1}^{N} \sum_{j=1}^{N} \vert \Pi(t)_{ij} \vert^2},
\end{equation}
where $N$ is the number of rows and columns in the transition matrices.

Using this norm we constructed a distance matrix $D$, whose elements are the distances between the matrices of the series $\Pi(t_0),\Pi(t_1),\ldots,\Pi(t_{T})$. In other words, for any $(i,j) \in  \llbracket 0,T \rrbracket^2 $ an element of $D$ reads:
\begin{equation}
D_{ij}=\Vert \Pi(t_i)- \Pi(t_j) \Vert.
\end{equation}
We then proceeded to perform an agglomerating clustering with a ward linkage method using the function sklearn.cluster.AgglomerativeClustering, available in the \texttt{sklearn} Python library~\cite{ScikitAgglomerative}. In this bottom-up algorithm, pairs of nodes and then pairs of clusters are recursively merged such that the variance of the distances within the clusters have, for each step, the least possible increase. 

\begin{figure}[!ht]
  \centering
 \includegraphics[width=\linewidth]{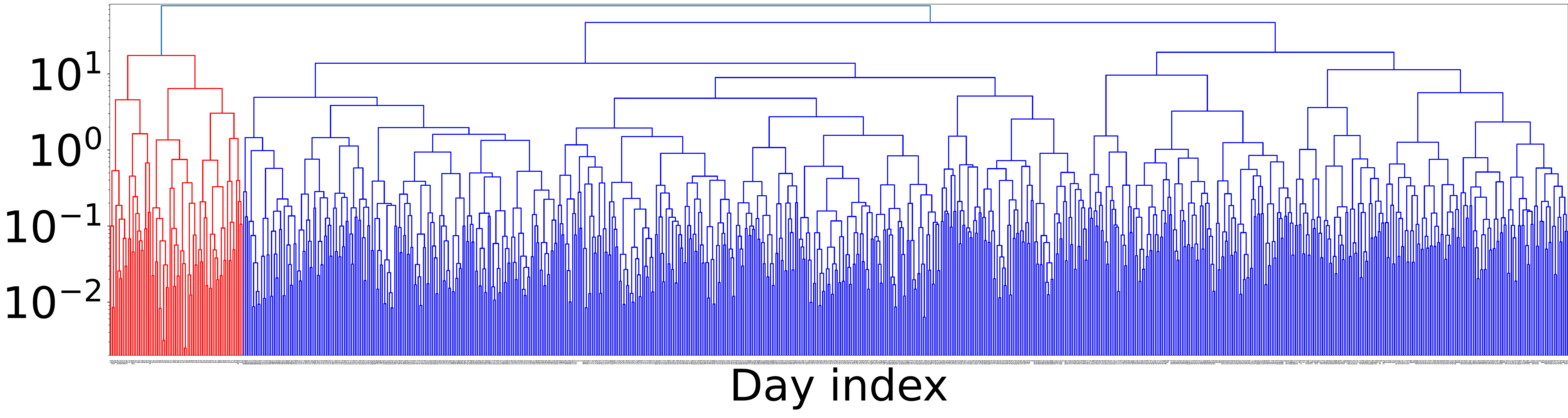}
 \includegraphics[width=\linewidth]{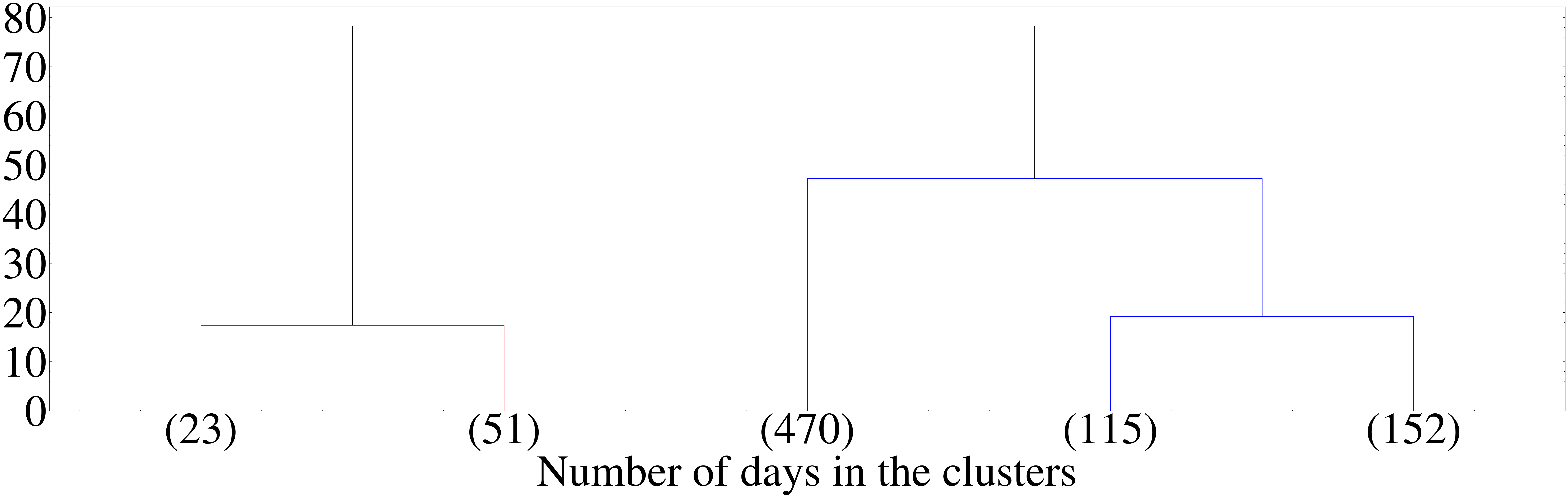}
 \caption{Hierarchical clustering dendrogram of the day-by-day transition matrices}
 \label{fig:dendrogram}
\end{figure}

The clustering process can be represented in a tree (dendrogram) in which the child branches at each step represent the pairs of clusters that merge into a parent branch. We report this hierarchical clustering dendrogram in Fig.~\ref{fig:dendrogram}. 
The length of the branches ($y$-axis) corresponds to the cophenetic distance, a distance which measures the level of similarity between two merged clusters~\cite{Copher}.\\

The top panel of \ref{fig:dendrogram} displays the full dendrogram from individual nodes to one unique cluster. On the bottom panel, this dendrogram is cut at the level of 5 clusters, after which the cophenetic distance increases significantly; the numbers in parentheses (in the $x$-axis) are the number of nodes belonging to each cluster.

\section{Spatial Clustering}

Spatial clustering into communities are obtained starting from the most representative matrices of the two main temporal clusters $C_0$ and $C_1$; these correspond to the unconfined and confined periods respectively, and are represented in  Fig.~\ref{fig:TemporalClustering}. 

\subsection*{Most representative current matrices}

We computed the mean matrices $\overline{\Pi}^{C_0}$ and $\overline{\Pi}^{C_1}$ and the matrices belonging to the unconfined ($C_0$) and confined ($C_1$) temporal clusters. From the mean transition matrices, we selected the most representative ones of each cluster by taking the daily (weekly rolled-average) transition matrix closest to the mean:
\begin{equation}
\widetilde{\Pi}^{C_k}=\min_{t\in{C_k}}{\Vert \Pi(t) - \overline{\Pi}^{C_k}\Vert },  k\in\{0,1\},   
\end{equation}
where $C_k$ is the set of days ${t_i}$ within the temporal cluster $k$.

The transition matrices defined above provide the daily probability of going from one province to another, but the weights do not contain any information on the population of each province. Hence, using the most representative transition matrices of the two principal temporal clusters and
the Istat vector $\boldsymbol{\rho}^{\text{Istat}}$, we defined the most representative current matrix $J^{C_k}$ as follow:
\begin{equation}
    J^{C_k}_{ij}= \widetilde{\Pi}^{C_k}_{ij} \boldsymbol{\rho}^{\text{Istat}}_i,\, k\in {0,1},
\end{equation}
subject to the normalization condition:
\begin{equation}
    \sum^N_{i,j} J^{C_k}_{ij}=1.
\end{equation}

We specify here that we do not define the current matrix using the stationary (Perron-Frobenius) population vector $\boldsymbol{\rho}^*$ but with the one computed from Istat data which is comparable up to a few fluctuation, as can be seen in Fig.~\ref{fig:CompareNodesPop_FB_Usage}. While this means that the detailed balance is not exactly verified, the detailed balance condition is not used in the clustering and  the population data of Istat is more accurate, thus ensuring that the computed currents are more representative of the real fluxes. 

\subsection*{Greedy Modularity Communities method (GMC)}\label{sec:MethodGreedyModularity}

The greedy modularity communities algorithm is provided by the \texttt{networkx} Python library (\texttt{greedy\_modularity\_communities}). This algorithm, developed in \cite{Newman2004_1} and refined in \cite{Newman2004_2, Newman2004_3}, relies on the optimization of the modularity $Q$. Let $W_{ij}$ be a weighted matrix, without self-loops, of the associated graph; for a given clustering $c$, the modularity is defined as~\cite{Newman2004_3}:
\begin{equation}\label{eq:modularity}
Q=\frac{1}{2m}\sum_{ij} W_{ij}- \frac{k_i k_j}{2m} \delta(c_i,c_j)    
\end{equation}
where $m=\frac{1}{2}\sum_{i,j} W_{ij}$  generalises what would be the number of edges in a binary graph, $k_i=\sum_j W_{ij}$ is the generalised degree of the node $i$, and $c_i$ labels the cluster to which node $i$ belongs.

To understand its meaning, consider the simpler case of an unweighted graph, where  $W_{ij}=A_{ij}$ is the adjacency matrix. If connections are made at random but respecting the degrees $k_i$ and $k_j$ of the nodes $i$ and $j$, then the probability of an existing link between these two nodes is $k_ik_j/2m$. This means that the modularity measures the difference between the linkage of the node within a community cluster and what is expected from a random network. With increasing values of $Q$, one has an increasing deviation from a random choice of linkage. Also, looking at Eq.\ref{eq:modularity}, we see that if there is only one cluster, then $\delta(c_i,c_j)\equiv 1$, and it is straightforward to see that in this case $Q=0$. In the opposite situation, where the clustering is made only of singleton then $\delta(c_i,c_j)=\delta_{ij}$; in this case as well, we see that $Q=0$. It is possible to show \cite{Brandes2008} that, in between these extreme cases, there exists an optimal clustering corresponding to maximal modularity.
The algorithm tests different levels of resolution through an agglomerative clustering method similar to the one presented in section~\ref{sec:TempClusMethods}, aiming at finding the clustering of the network with maximal modularity. 

\subsection*{Effective distance matrix between nodes}
Following ref~\cite{Brockmann2013}, we define the effective distance between two adjacent nodes $i$ and $j$ as:
\begin{equation}
d_{ij}=1-\ln{\Pi_{ij}}.
\end{equation}
If there exists a path going from $i$ to $j$ with $l$ steps $$\Gamma_{ij}=\{(k_0=i,k_1),(k_1,k_2),\ldots,(k_{l-1},k_{l}=j)\},$$ the direct length of a path is the sum of the effective distances along its steps: 
\begin{equation}
\lambda(\Gamma_{ij})=\sum_{n=0}^{l-1} d_{k_n,k_{n+1}} .
\end{equation}
We defined the effective distance as the minimal distance among all the existing paths from $i$ to $j$:
\begin{equation}
D_{ij}=\min_{\Gamma_{ij}}\lambda(\Gamma_{ij})    
\end{equation}
Then for any two nodes $i$, $j$ of the network defined by $\Pi$, the effective distance matrix is the symmetric part,
\begin{equation}
\Delta^S=(\Delta+\Delta^t)/2 ,
\end{equation} of the matrix $\Delta$, whose elements are defined as follows: 

\begin{equation}\label{eq:EffectiveDistance}
\Delta_{ij} =\left\{
	\begin{array}{ll}
		0      & \mbox{if }\, i=j \\
		d_{ij} & \mbox{if }\, \Pi_{ij}\neq 0\\
        d_{ji} & \mbox{if }\, \Pi_{ij}=0 \, \mbox{and}\, \Pi_{ji}\neq 0\\
        D_{ij} & \mbox{if }\, \exists \, \Gamma_{ij}\\
        D_{ji} & \mbox{if }\,  \nexists\, \Gamma_{ij} \, \mbox{and}\, \exists \, \Gamma_{ji} \\
        +\infty& \mbox{elsewhere.}
	\end{array}
\right.
\end{equation}

This definition is valid for any weighted directed graph. In particular, the last line is not needed if the graph is weakly connected ($\forall (i,j),\, \exists\, \Gamma_{ij}\, \text{or}\, \exists\, \Gamma_{ji} $). Similarly, the two last lines are not needed if it is strongly connected ($\forall (i,j),\, \exists\, \Gamma_{ij}$).\\
In our case, the most representative transition matrix of the non-confined period, $\overline{\Pi}^{C_0}$ is strongly connected while $\overline{\Pi}^{C_1}$, the graph associated with the most representative transition matrix for the confinement period is not even weakly connected, and its connected components are not always strongly connected.
\\
We add that, on a computer,  `infinite' must be represented as a large number; this value was defined as $100$ times the maximum of the well-defined elements of $\Delta$. The effective distance matrix was normalized by its mean value: $\Delta^S \leftarrow \Delta^S / \overline{\Delta^S}$ where $\overline{\Delta^S}=\frac{1}{N^2}\sum^N_{i,j} \Delta^S_{ij}$. In this way, the agglomerative clustering operations on the distance matrix do not depend on the large-scale cutoff.

\subsection*{Critical Variable Selection method (CVS)}\label{sec:MethodResRel}

The resolution-relevance method~\cite{Obregon2018,Cubero2018,Cubero2019,Cubero2020,Marsili2022,Holtzman2022} 
has been successful in identifying optimal clustering for the reduction of complexity in the representation of biomolecules \cite{Giulini2020} or for a protein conformational landscape \cite{Mele2022}.

Considering a set of $N$ objects and a given clustering of them, we labeled the $K$ clusters by $s\in \llbracket 1, K \rrbracket $ and defined $k_s$ to be the number of objects in cluster $s$. $k_s/N$ is the empirical probability for an object to belong to cluster $s$. The \emph{resolution} is defined as the Shannon entropy of this probability distribution:
\begin{equation}
H[s]= -\sum^{K}_{s=1} \frac{k_s}{N} \log_N{\frac{k_s}{N}} 
\end{equation}
where $\log_N$ is the logarithm in base $N$ such that $\log_N N = 1$. $H[s] = 0$ when all objects belong to only one cluster, and $H[s]=1$ at the other extreme, when each object has its own separate cluster.

Resolution alone, however, is not sufficient to identify an optimal level of informativeness of a given clustering. A second quantity, the relevance $H[k]$, is defined based on the number of clusters containing $k$ objects, $m_k$~\cite{Marsili2013}:
\begin{equation}
m_k=\sum^{K}_{s=1}\delta_{k,k_s}.
\end{equation}
The \emph{relevance} is defined as follows:
\begin{equation}
H[k]=-\sum^{N}_{k=1} \frac{km_k}{N} \log_N{\frac{km_k}{N}}.
\end{equation}

In the latter expression, the factor $\frac{km_k}{N}$ is the empirical probability that a randomly chosen object in the collection belongs to the cluster with $k$ elements in it. The relevance is the Shannon entropy associated with this second empirical probability.

For both limit cases of $1$ and $N$ clusters, $H [k] = 0$, the relevance being non-negative otherwise~\cite{Marsili2013, Mele2022}. The maximum relevance thus corresponds to an optimal clustering, i.e. to the most informative partition of the collection of objects.

\begin{figure}[ht]
    \centering
       \includegraphics[width=0.75\linewidth]{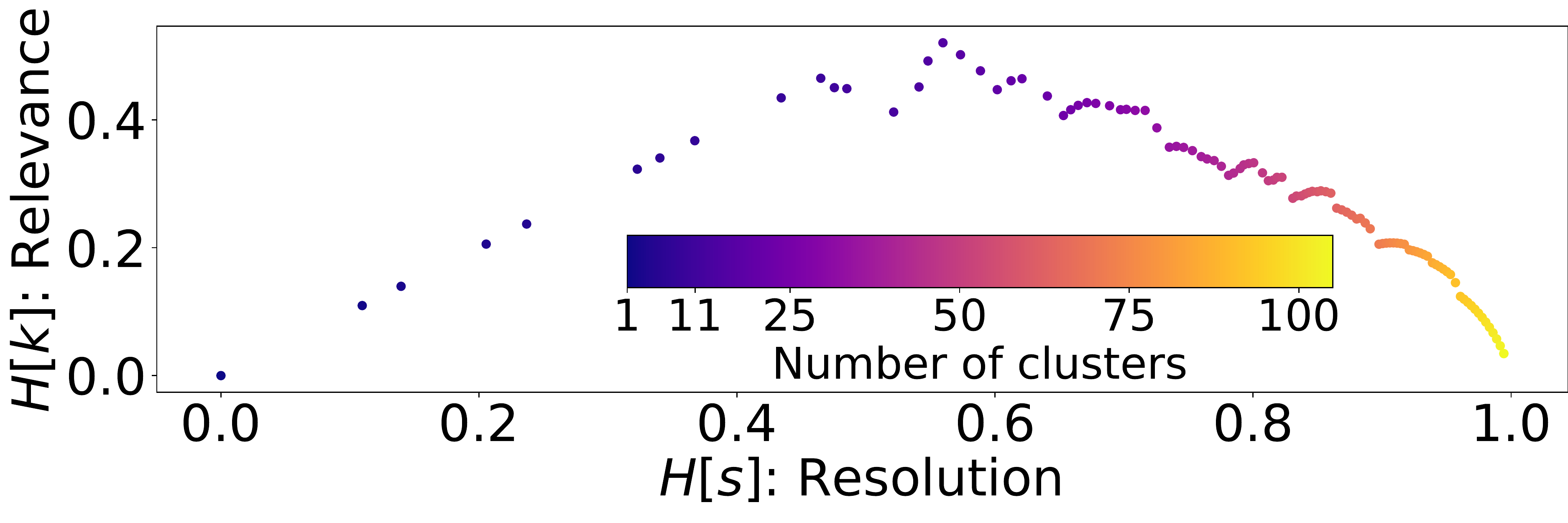}
       \includegraphics[width=0.75\linewidth]{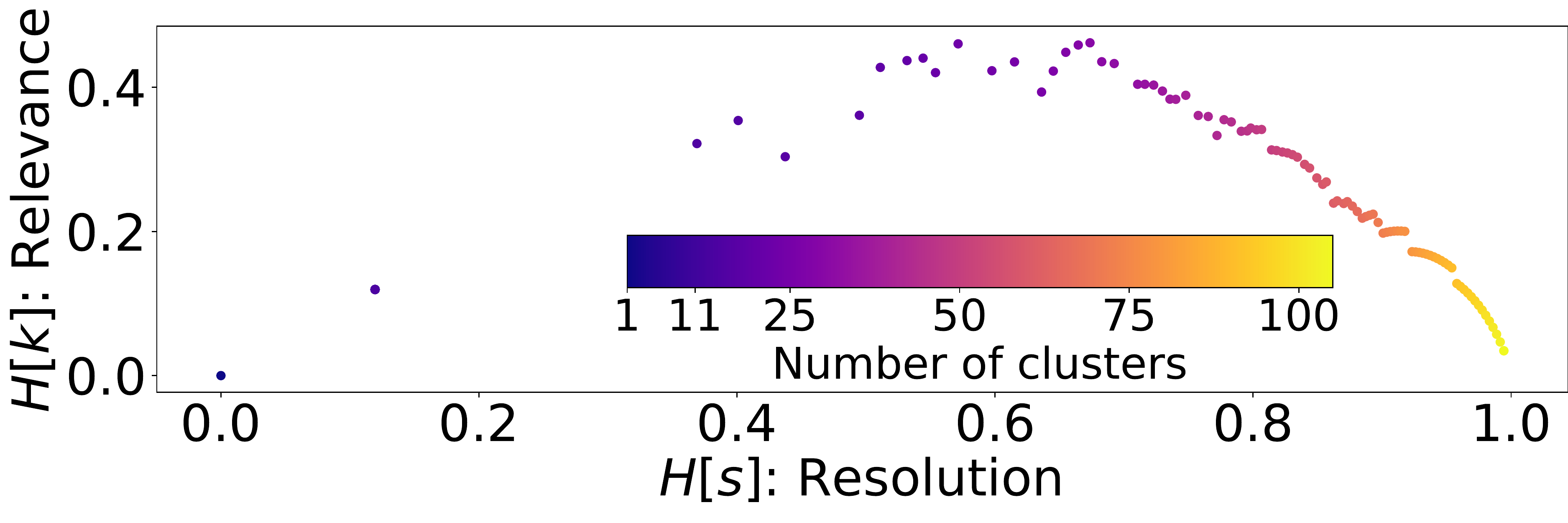}
    \caption{Resolution versus relevance for agglomerative spatial clustering of temporal cluster $C_0$ (confined, top) and $C_1$ (unconfined, bottom).}
    \label{fig:Res_RelC0}
\end{figure}

We performed an agglomerative clustering of the nodes representing provinces using the distance introduced above, and computed for each number of clusters from $1$ to $N$ the corresponding values of resolution and relevance (see Fig.~\ref{fig:Res_RelC0} ). The optimal partition of provinces was defined as the clustering with the maximum relevance value. 
\let\subsection\section
\let\subsubsection\subsection
\bibliographystyle{apsrev4-2}
\bibliography{Epidemic-Network-Inference,Epidemic-Italy-Websites}

\onecolumngrid
\sifigures
\newpage
\appendix
\section{Supplementary Information}
\let\section\subsection
\let\subsection\subsubsection
\let\tiny\footnotesize
\section{List of Provinces used in the study}
\begin{tiny}
\begin{tabular}{llllr}\label{table:provinces}
{node} &                      Provinces & Car plate  &    ITTER107     & Population \\
{index} &                            name   &          code &       code& 01 Jan 2020 \\
\hline
0   &              Agrigento &          AG &        ITG14 &              412427.0 \\
1   &            Alessandria &          AL &        ITC18 &              407049.0 \\
2   &                 Ancona &          AN &        ITE32 &              461745.0 \\
3   &                  Aosta &          AO &        ITC20 &              123337.0 \\
4   &          Ascoli Piceno &          AP &        ITE34 &              202317.0 \\
5   &               L'Aquila &          AQ &        ITF11 &              288439.0 \\
6   &                 Arezzo &          AR &        ITE18 &              334634.0 \\
7   &                   Asti &          AT &        ITC17 &              207939.0 \\
8   &               Avellino &          AV &        ITF34 &              399623.0 \\
9   &                   Bari &          BA &        ITF42 &             1224756.0 \\
10  &                Bergamo &          BG &        ITC46 &             1102670.0 \\
11  &                 Biella &          BI &        ITC13 &              169560.0 \\
12  &                Belluno &          BL &        ITD33 &              198518.0 \\
13  &              Benevento &          BN &        ITF32 &              263460.0 \\
14  &                Bologna &          BO &        ITD55 &             1015701.0 \\
15  &               Brindisi &          BR &        ITF44 &              379851.0 \\
16  &                Brescia &          BS &        ITC47 &             1254322.0 \\
17  &  Barletta-Andria-Trani &          BT &        IT110 &              379251.0 \\
18  &                Bolzano &          BZ &        ITD10 &              535774.0 \\
19  &  Cagliari|Sud Sardegna &          CA &  ITG27|IT111 &              754878.0 \\
20  &             Campobasso &          CB &        ITF22 &              210599.0 \\
21  &                Caserta &          CE &        ITF31 &              900293.0 \\
22  &                 Chieti &          CH &        ITF14 &              372473.0 \\
23  &          Caltanissetta &          CL &        ITG15 &              250550.0 \\
24  &                  Cuneo &          CN &        ITC16 &              580789.0 \\
25  &                   Como &          CO &        ITC42 &              594657.0 \\
26  &                Cremona &          CR &        ITC4A &              351287.0 \\
27  &                Cosenza &          CS &        ITF61 &              671171.0 \\
28  &                Catania &          CT &        ITG17 &             1068835.0 \\
29  &              Catanzaro &          CZ &        ITF63 &              341991.0 \\
30  &                   Enna &          EN &        ITG16 &              155982.0 \\
31  &           Forlì-Cesena &          FC &        ITD58 &              391524.0 \\
32  &                Ferrara &          FE &        ITD56 &              340755.0 \\
33  &                 Foggia &          FG &        ITF41 &              597902.0 \\
34  &                Firenze &          FI &        ITE14 &              994717.0 \\
35  &                  Fermo &          FM &        IT109 &              168485.0 \\
36  &              Frosinone &          FR &        ITE45 &              468438.0 \\
37  &                 Genova &          GE &        ITC33 &              816250.0 \\
38  &                Gorizia &          GO &        ITD43 &              138666.0 \\
39  &               Grosseto &          GR &        ITE1A &              216989.0 \\
40  &                Imperia &          IM &        ITC31 &              208561.0 \\
41  &                Isernia &          IS &        ITF21 &               80170.0 \\
42  &                Crotone &          KR &        ITF62 &              161744.0 \\
43  &                  Lecco &          LC &        ITC43 &              332435.0 \\
44  &                  Lecce &          LE &        ITF45 &              772276.0 \\
45  &                Livorno &          LI &        ITE16 &              326716.0 \\
46  &                   Lodi &          LO &        ITC49 &              227064.0 \\
47  &                 Latina &          LT &        ITE44 &              565840.0 \\
48  &                  Lucca &          LU &        ITE12 &              381890.0 \\
49  &        Monza e Brianza &          MB &        IT108 &              870112.0 \\
50  &               Macerata &          MC &        ITE33 &              305249.0 \\
51  &                Messina &          ME &        ITG13 &              599990.0 \\
52  &                 Milano &          MI &        ITC45 &             3237101.0 \\
\hline
\end{tabular}
\end{tiny}
\begin{tiny}
\begin{tabular}{llllr}
{node} &                      Provinces & Car plate  &    ITTER107     & Population \\
{index} &                            name   &          code &       code& 01 Jan 2020 \\
\hline
53  &                Mantova &          MN &        ITC4B &              404440.0 \\
54  &                 Modena &          MO &        ITD54 &              702787.0 \\
55  &          Massa-Carrara &          MS &        ITE11 &              188395.0 \\
56  &                 Matera &          MT &        ITF52 &              191663.0 \\
57  &                 Napoli &          NA &        ITF33 &             2967117.0 \\
58  &                 Novara &          NO &        ITC15 &              361845.0 \\
59  &        Nuoro|Ogliastra &          NU &        ITG26 &              199349.0 \\
60  &               Oristano &          OR &        ITG28 &              150812.0 \\
61  &                Palermo &          PA &        ITG12 &             1199626.0 \\
62  &               Piacenza &          PC &        ITD51 &              283889.0 \\
63  &                 Padova &          PD &        ITD36 &              930898.0 \\
64  &                Pescara &          PE &        ITF13 &              313346.0 \\
65  &                Perugia &          PG &        ITE21 &              641318.0 \\
66  &                   Pisa &          PI &        ITE17 &              417245.0 \\
67  &              Pordenone &          PN &        ITD41 &              310158.0 \\
68  &                  Prato &          PO &        ITE15 &              264397.0 \\
69  &                  Parma &          PR &        ITD52 &              450044.0 \\
70  &                Pistoia &          PT &        ITE13 &              289256.0 \\
71  &        Pesaro e Urbino &          PU &        ITE31 &              351993.0 \\
72  &                  Pavia &          PV &        ITC48 &              534691.0 \\
73  &                Potenza &          PZ &        ITF51 &              348336.0 \\
74  &                Ravenna &          RA &        ITD57 &              386007.0 \\
75  &     Reggio di Calabria &          RC &        ITF65 &              518978.0 \\
76  &     Reggio nell'Emilia &          RE &        ITD53 &              524193.0 \\
77  &                 Ragusa &          RG &        ITG18 &              315082.0 \\
78  &                  Rieti &          RI &        ITE42 &              150689.0 \\
79  &                   Roma &          RM &        ITE43 &             4222631.0 \\
80  &                 Rimini &          RN &        ITD59 &              336916.0 \\
81  &                 Rovigo &          RO &        ITD37 &              229097.0 \\
82  &                Salerno &          SA &        ITF35 &             1060188.0 \\
83  &                  Siena &          SI &        ITE19 &              262046.0 \\
84  &                Sondrio &          SO &        ITC44 &              178208.0 \\
85  &              La Spezia &          SP &        ITC34 &              214879.0 \\
86  &               Siracusa &          SR &        ITG19 &              383743.0 \\
87  &   Sassari|Olbia-Tempio &          SS &        ITG25 &              474142.0 \\
88  &                 Savona &          SV &        ITC32 &              267748.0 \\
89  &                Taranto &          TA &        ITF43 &              558130.0 \\
90  &                 Teramo &          TE &        ITF12 &              299402.0 \\
91  &                 Trento &          TN &        ITD20 &              542158.0 \\
92  &                 Torino &          TO &        ITC11 &             2205104.0 \\
93  &                Trapani &          TP &        ITG11 &              415233.0 \\
94  &                  Terni &          TR &        ITE22 &              218254.0 \\
95  &                Trieste &          TS &        ITD44 &              230623.0 \\
96  &                Treviso &          TV &        ITD34 &              876755.0 \\
97  &                  Udine &          UD &        ITD42 &              517848.0 \\
98  &                 Varese &          VA &        ITC41 &              878059.0 \\
99  &   Verbano-Cusio-Ossola &          VB &        ITC14 &              154233.0 \\
100 &               Vercelli &          VC &        ITC12 &              165760.0 \\
101 &                Venezia &          VE &        ITD35 &              839396.0 \\
102 &                Vicenza &          VI &        ITD32 &              852861.0 \\
103 &                 Verona &          VR &        ITD31 &              927108.0 \\
104 &                Viterbo &          VT &        ITE41 &              307592.0 \\
105 &          Vibo Valentia &          VV &        ITF64 &              150702.0 \\
\hline
\end{tabular}
\end{tiny}

\begin{figure}[H]
    \centering
    \includegraphics[width=0.7\linewidth]{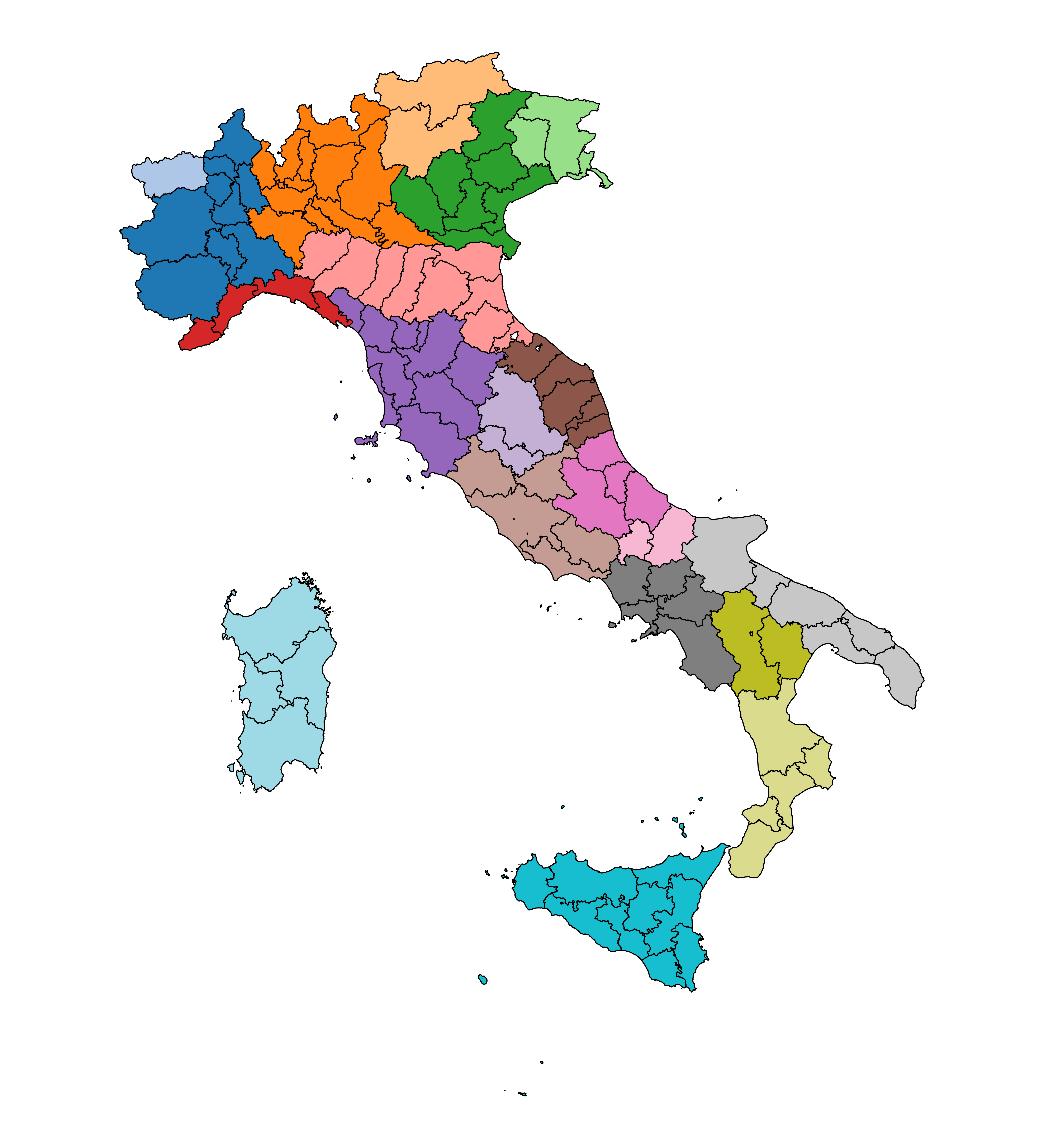}
    \caption{Administrative borders of the provinces of Italy considered in the study (black lines) and of the regions (in colors) as there are defined now.}
    \label{fig:RegionAndPorvItaly}
\end{figure}
\begin{figure}[H]
    \centering
    \includegraphics[width=\linewidth]{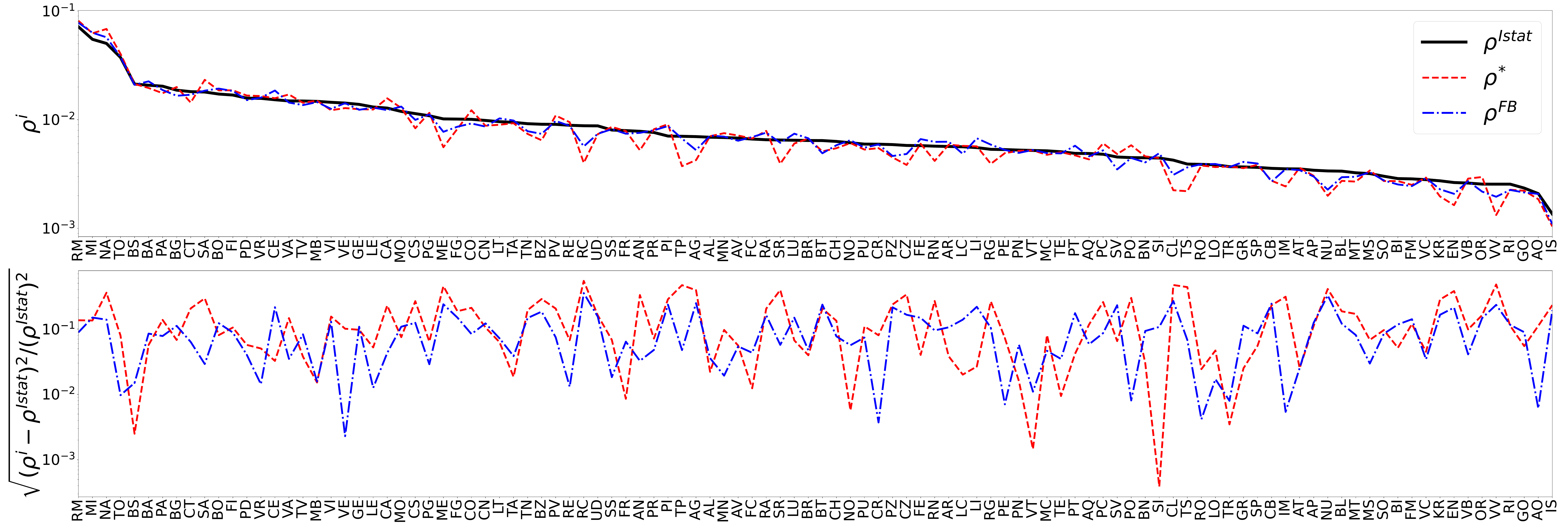}
    \caption{Top Panel: Comparison of the different population density vector from Facebook and Istat data: $\boldsymbol{\rho}^{\text{Istat}}$, $\boldsymbol{\rho}^{\text{FB}}$ and $\boldsymbol{\rho}^{*}$ against provinces order by Istat population.
    Bottom Panel: Standard deviation of the Facebook data vectors from the Istat data vector is displayed showing a good agreement between datasets.}
    \label{fig:CompareNodesPopAllregions}
\end{figure}
\section{Region of Italy organized by Emperor Augustus}


\begin{figure}[H]
    \centering
    \includegraphics[height=6.3cm]{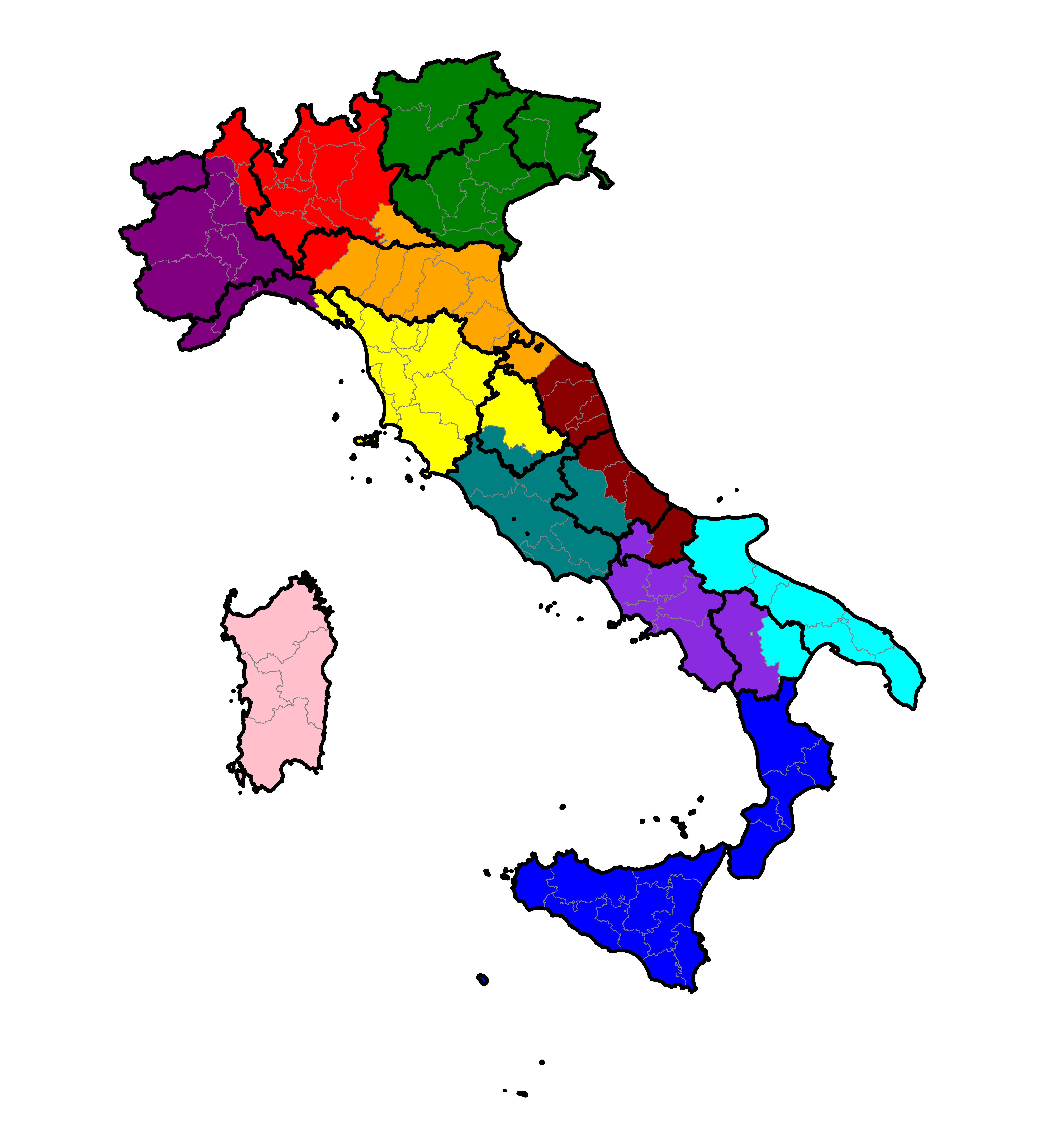}
    \includegraphics[height=6.3cm]{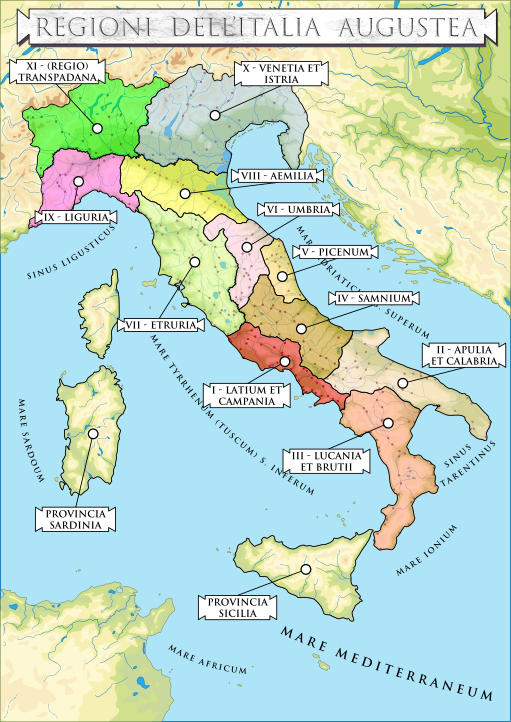}
    \includegraphics[height=6.3cm]{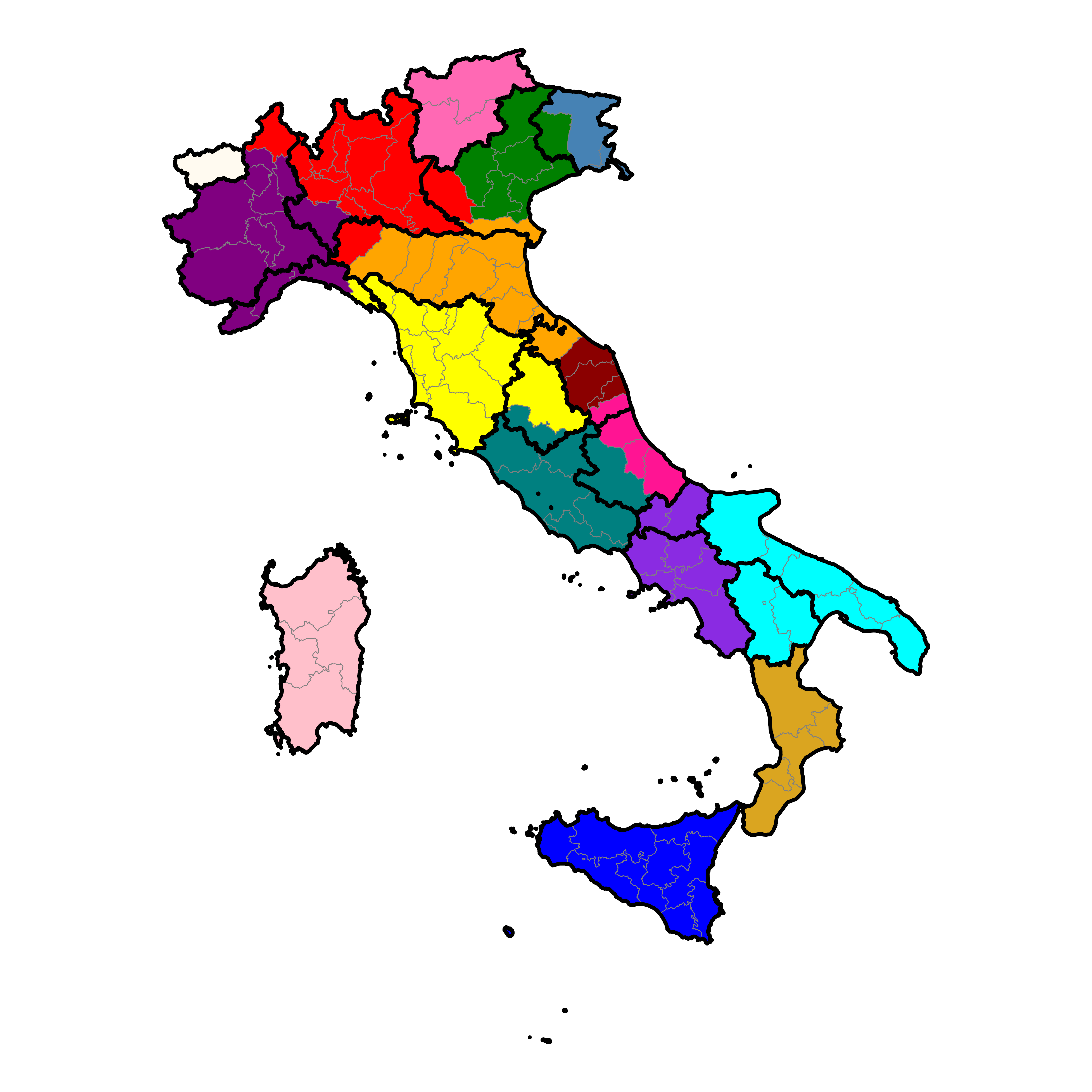}
    \caption{The proposed partition of Emperor Augustus, together with the optimal spatial clustering for the non-confided periods (left with GMC, right with CVS). Source central map : \href{https://it.wikipedia.org/wiki/Regioni_dell\%27Italia\_augustea\#/media/File:Regioni\_dell'Italia\_Augustea.svg}{https://it.wikipedia.org/wiki/Regioni\_dell\%27Italia\_augustea\#/media/File:Regioni\_dell'Italia\_Augustea.svg} 
    }
\end{figure}

\pagebreak

\section{Data Preparation}\label{data-preparation}

\subsection{Data Sources}\label{data-sources}

Mainly two data sources were used for our analysis: the \textbf{Facebook
Data for Good Italy Coronavirus Disease Prevention Map} data and the
\textbf{COVID-19 data} published by the ISS (Istituto Superiore di
Sanità - National Institute of Health) - INFN (Istituto Nazionale di
Fisica Nucleare - National Institute for Nuclear Physics) collaboration.

Respectively, the two data sources contained the following datasets that
were of interest to us:

\begin{itemize}
\item
  The \textbf{Movements Between Administrative Regions} dataset
  describes the number of Facebook users that move between two NUTS-3
  administrative regions (aka \emph{province}). The temporal aggregation
  of the dataset is of 8 hours, meaning that if a person is
  checked-in in region A in a certain time frame, and the same person is
  found to be checked-in in another region B in the subsequent time
  frame, then a movement between regions A and B are counted. A 24-hour
  the day is divided into three time frames: 00:00-08:00, 08:00-16:00 and
  16:00-24:00. Two types of data are considered: the baseline, which is
  computed by taking the average on the same weekday for the same
  weekdays, and the people during the crisis, which is the actual number of
  people detected in the specified DateTime. Only users of the Facebook
  app that have the Location History option enabled are counted, and
  also if the aggregation yields counts under 10 units then the datum
  is discarded.
\item
  The \textbf{New Positive Cases By Date} dataset describes the number
  of new positive SARS-CoV-2 cases, aggregated by date and province of
  detection. This dataset does not suffer from the lag between detection
  and publication, unlike the Dipartimento della Protezione Civile
  (Department for Civil Defense) data. The number of cases is the result
  of a window average over a week, where the final result is the day in
  the middle of the week (the fourth day of the week).
\end{itemize}

\subsection{The choice of the stack}\label{the-choice-of-the-stack}

In order to perform the extraction, loading, and transformation of the
data various paths have been explored, but our choice fell on the
current technological stack.

\begin{itemize}
\item
  \textbf{Python} is the main scripting language and piece of software
  used throughout the whole pipeline. Its user-friendliness, its
  widespread use among both the industry and researchers, and the
  availability of great libraries for data science and visualization
  made it our natural choice for our purposes. In particular, the
  libraries mainly used by us are Selenium (web browser automation) and
  Pandas (data analysis and manipulation).
\item
  \textbf{Miller} is a toolkit for data munging. It allows quick CSV
  manipulations and it contains several powerful commands, that can also
  be chained one after the other.
\item
  \textbf{Bash} is used for integrating and preprocessing various data
  sources. It is extremely flexible and compatible with most of
  Unix-like systems, and for certain types of data science workloads it
  can quickly and efficiently get the work done.
\item
  \textbf{DuckDB} is an embeddable analytical database. It is similar
  to SQLite, in that the database system runs within a host process, but
  it is optimized for analytical (OLAP) workloads. It allows for
  manipulations \emph{à la} Pandas but also has full-query optimization
  and transactional storage. It is a good choice for our purposes since
  it allows faster queries to be made, it does not require the
  maintenance of a DB stack and it integrates very well with Python
  thanks to the DuckDB Python API.
\end{itemize}

\subsection{Gathering the data}\label{gathering-the-data}

The \textbf{Facebook Data for Good} data can only be downloaded by using
an online interface, and each transaction is size-capped (i.e.~Movements
Between Administrative Regions data for more than a two-weeks period
would not be downloaded). In order to facilitate and speed up the
sourcing of the data, a bulk download tool was developed. The tool makes
use of Selenium, a Python library for browser automation, that allows automated
workflows that simulate human interaction with a browser. The resulting
raw data are available as zip archives containing many CSV files.
\\
The \textbf{COVID-19 data} by ISS/INFN is published as a single zip
archive containing many CSV files, one for each aggregation, data type
and province/region.
\\
Reference tables are also essential for the analysis, as they allow data
integration between incoherent definitions (in particular, concerning
spatial aggregation units) between different datasets. Some of them are
manually compiled, and others are aggregate data extracted from the original
datasets:

\begin{itemize}
\item
  The \textbf{provinces} identifications conversion table has been
  created manually. It contains the correspondence between IDs for
  provinces in the Facebook dataset, the ISS/INFN dataset, and the ``car
  number plate code'' two letters characters).
\item
  The \textbf{locations} reference table contains the latitude and
  longitude for each province.
\end{itemize}

\subsection{Cleaning and loading the
data}\label{cleaning-and-loading-the-data}

The raw data is then loaded into our database for further analysis.

First of all, the archives are unpacked and the data is cleaned for our
purposes with the use of Miller. The following operations are performed:

\begin{itemize}
\item
  The data is filtered in order to get only data for Italian provinces
  (generally Facebook uses rectangular bounding boxes to get subsets of
  data).
\item
  Minor changes in data formats are operated (such as missing date-time
  imputation and format correction for date-time strings).
\item
  Null entries are discarded.
\item
  Only columns of our interest are selected.
\end{itemize}
Then the data are piped through an SQL COPY command, that loads it in a
DuckDB database.

\subsection{Transforming the data}\label{transforming-the-data}

This is the last step of our data preparation pipeline.

In this step, we transform the raw data contained in the database into SQL
tables with SQL views, in such a way that they can be accessed easily
and are expressed in a manner that is optimal for the analysis purposes
of our research.

The Movements Between Administrative Regions dataset has rows aggregated
and summed over with the new definitions of provinces as defined in the
reference table.

Starting from this table, then multiple views are created:

\begin{itemize}
\item total number of people moving from each origin place by date;
\item total number of people moving between places summed over by each day;
\item daily probability that a movement between places happens (aka \emph{transition matrix});
\item total probability that a movement between places happens;
\item weekly rolling average of the daily probability of movement between
  places.
\end{itemize}

The COVID-19 ISS row dataset is also aggregated and summed
over the province using the new definitions as defined in the reference
table.\ref{table:provinces}. \\

\section{About Perron-Frobenius (PF) theorem for stochastic matrix}

In the graph identified by the mean matrix $\overline{\Pi}$ there is a non-zero probability to reach any node from any other node in a finite number of steps, that is, the graph is strongly connected and aperiodic. (To say a graph is aperiodic is equivalent to saying that its representative matrix is irreducible or saying that the random walk on the graph is ergodic.) Then, the transition matrix representing the graph is non-negative and irreducible. 

For a general non-negative irreducible matrix, $\boldsymbol{\Pi}$, the PF theorem then ensures that the highest eigenvalue $\lambda^*$ of $\boldsymbol{\Pi}$ is not degenerate. It is often called the PF eigenvalue and we name PF left eigenvector $l^*$ (and right eigenvector $r^*$) its associated eigenvectors.

A consequence of the theorem in this case is also that any distribution on which we apply the matrix successively, will concentrate, in the long time (long path limit), to the stationary density vector $\rho^*_i=l^*_i.r^*_i$.

We explain it here in our specific case where the matrix is stochastic.

The normalization of $\boldsymbol{\Pi}$ is such that it is stochastic, i.e. each its raws sum to 1, i.e $1$ is eigenvalue of $\overline{\Pi}$ and is associated with the right eigenvector $\boldsymbol{r}^*=\mathbf{1}_{np}=(1,1,\ldots,1)^T$ and $\boldsymbol{l}^*$ :
\begin{equation}
\boldsymbol{l}^*=\boldsymbol{l}^*\overline{\Pi}
\end{equation}
Moreover, one can also show that in this case, $1$ is the maximum eigenvalue possible and the PF theorem ensure it is unique, as well as its associated eigenvectors. 

Therefore, for stochastic matrix, we commonly identify $\boldsymbol{\rho}^*=\boldsymbol{l}^*$ as the stationary density vector but in general, $r^*_i$ may be something else than the $\mathbf{1}$ and the PF eigenvalue different than 1.

In the following, we show the long-time limit convergence in our case.

Because the PF left eigenvector of $\boldsymbol{\Pi}$ is the unique invariant, it ensures the detail balance of the associated Markov process:
$$\rho^*_i \Pi_{ij}= \Pi_{ji} \rho^*_j$$.
Multiplying the left and the right by $\rho_i^{*-1/2}$ and $\rho_j^{*-1/2}$, we obtain that
\begin{equation}
\rho_i^{*1/2} \Pi_{ij}  \rho_j^{*-1/2} =  \rho_i^{*-1/2} \Pi_{ji} \rho_j^{*1/2} = (\rho_j^{*-1/2} \Pi_{ij} \rho_i^{*1/2})^T.
\end{equation}

Hence the matrix $S$ defined by the elements
\begin{equation}
    S_{ij}= \rho_i^{*1/2} \Pi_{ij}  \rho_j^{*-1/2},
\end{equation}
is equal to its own transposed, and so is symmetric.
\\
Defining the transformation $U=diag(\rho_i^{*1/2})$ we have:
\begin{equation}
    S=U^{-1} \boldsymbol{\Pi} U ,
\end{equation}
then $S$ have the same eigenvalues (for $S$, left and right eigenvector are the same) and is symmetric, so diagonalizable in real space, so $\Pi$ itself is diagonalizable.

So there exists a mapping, i.e. a base of the vector space, $O$ such that :
$$\Pi=O^{-1} \boldsymbol{\Lambda} O$$ with $\boldsymbol{\Lambda}=diag(\lambda^*,\lambda_1,....,\lambda_N)$.\\

The PF theorem ensures that for our stochastic matrix $\lambda^*=1>|\lambda_i|\ge 0,\, i=1,..,N$,\\
Hence, in the long time limit (or long path limit). 
$$\boldsymbol{\Lambda}^t \sim diag(1,0,0,0,...,0) $$
The principal (or Perron-Frobenius) eigenvalue $\lambda^*=1$ will dominate and any non-trivial distribution $\boldsymbol{\rho}$ over the nodes will converge to the unique stationary density vector:
$$\boldsymbol{\rho} \boldsymbol{\Pi}^t \sim \boldsymbol{\rho}^* .$$

\section{Clustering of the mean current Matrix}
\begin{figure}[H]
    \centering
    \includegraphics[width=0.6\linewidth]{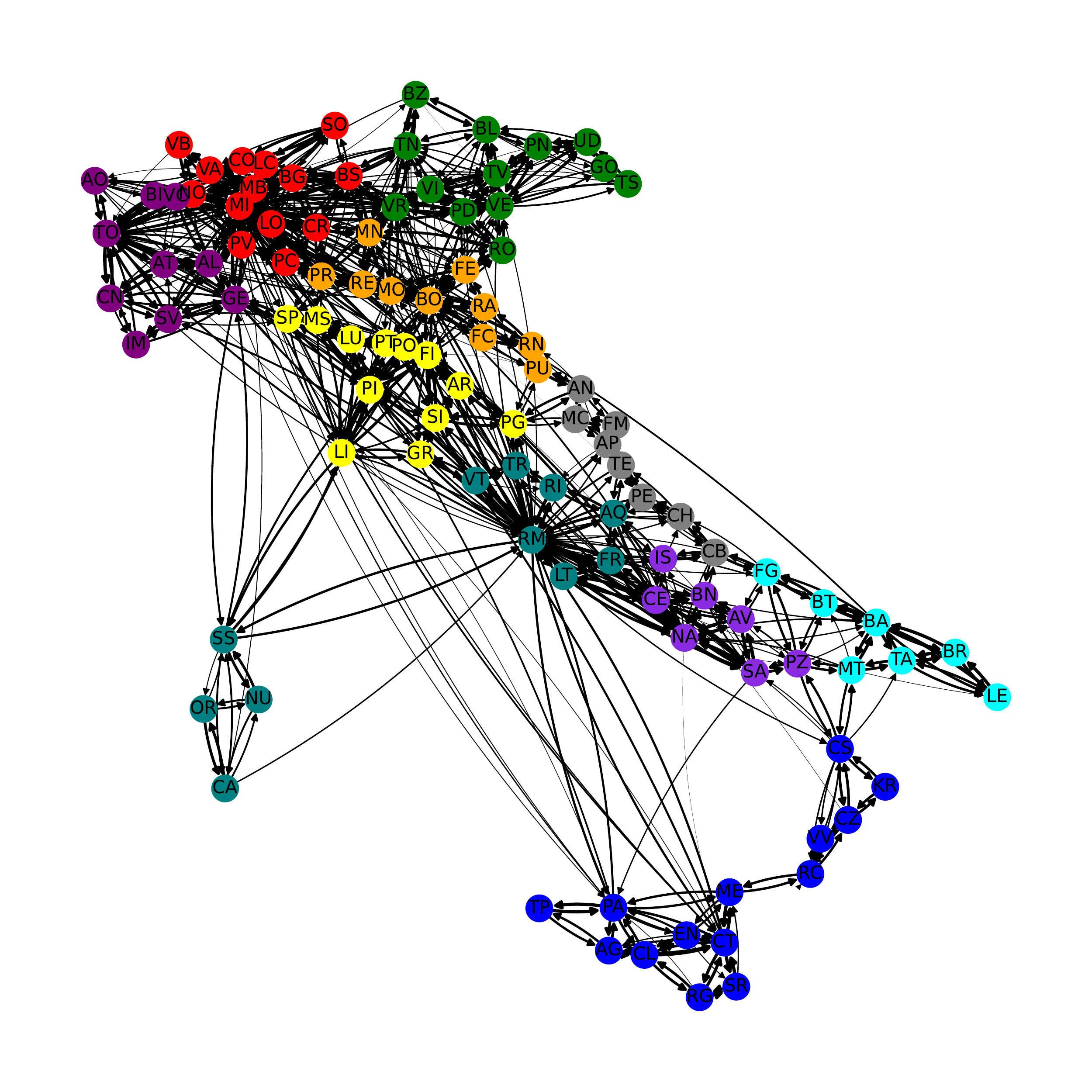}\\
    \includegraphics[width=0.6\linewidth]{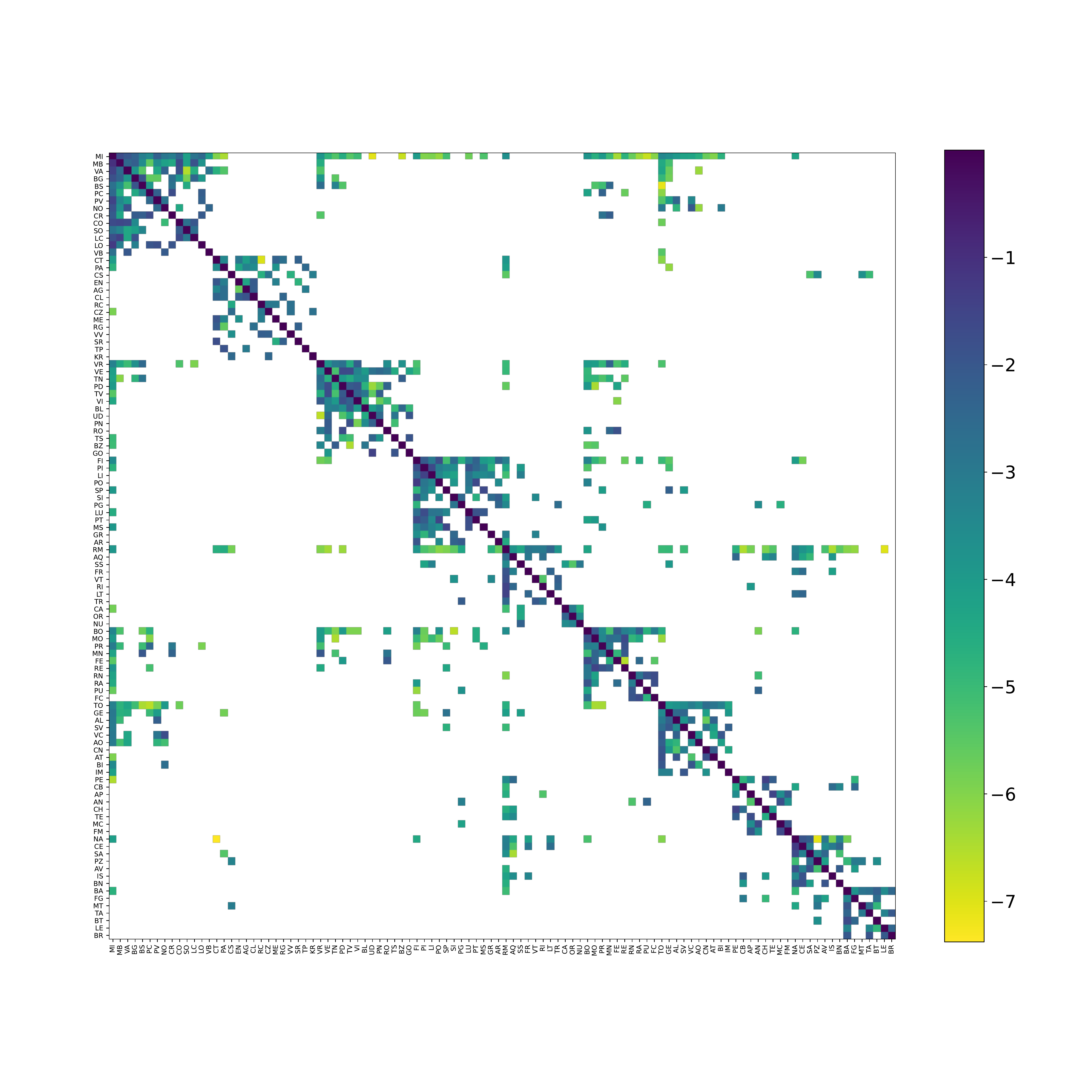}\\
    \caption{Italian provinces clustered using GMC by communities using the all time-averaged matrix.\\
    Top panel: Representation of the clustered mean current matrix, for visualization the shades are in log10 of the mean probabilities of going from one province to another.\\
    Bottom: Graph representation of the community clustering with colors corresponding to the different clusters, the widths of the links are proportional to the logarithm of the transition probability. At the exception of Sardinia that is in Rome cluster here, the clustering is the same than non-confined most representative matrix using GMC.}
    \label{fig:CommunitiesItalie}
\end{figure}

In Fig.\ref{fig:CommunitiesItalie}, on the top is displayed a representation of the mean current matrix sorted by clusters and by weight, one sees that the method is satisfactory giving well-defined blocks corresponding to each community.  On the bottom panel, we see that the clusters correspond, apart from very few border cases,(and Umbria is split apart) to a group of regions in Italy. In detail, for the ten clusters found, we have:
\begin{itemize}
 \item The green cluster corresponds perfectly to the ``Triveneto'' region (that is, Veneto, Friuli-Venezia-Giulia, and the provinces of Trento and Bolzano). 
 \item The red one to Lombardia with the exception of Mantova plus the two provinces of  Verbano/Cusio/Ossola  and Novara (belonging to Piemonte) and Piacenza (belonging to Emilia-Romagna).
 \item The dark purple corresponds to the region of Valle d'Aosta and Piemonte (minus VB and NO) and Liguria, at the exception of Spezia.
 \item The yellow cluster is Toscana plus the provinces of Spezia (Liguria) and Perugia (Umbria)
 \item The orange one corresponds to the region of Emilia-Romagna at the exception of Piacenza (PC) and adds the provinces of Pesaro/Urbino (Marche) and Mantova (Lombardia).
 \item The grey cluster is the regions of Marche (minus PU), Abruzzo (minus AQ) and Molise (minus IS)
 \item The teal one matches with the regions of Lazio and Sardegna (plus TE (Umbria) and AQ (Abruzzo))
 \item The light purple corresponds to the region of Campania plus the province of Isernia belonging to Molise.
 \item The light blue cluster corresponds perfectly o the regions of Puglia and Basilicata 
 \item Finally, the blue cluster perfectly to the regions of Calabria and Sicilia.
\end{itemize}
\pagebreak

\section{Z-score of the probability of going in and out of provinces}
To better see which provinces differ the most from the mean trend we compute the $Z$-score, which is defined for the time series $X_i(t)$ of province $i$ as:
\begin{equation}\label{eq:Zscore}
Z_i=\frac{1}{T}\sum_{t=0}^T \frac{|X_i(t)-\mu(t)|}{\sigma(t)}
\end{equation}
with $\mu(t)=\left<X_i(t)\right>$ being the average over provinces and $\sigma(t)=\sqrt{\left<X_i(t)-\mu(t)\right>}$ the standard deviation at time $t$.

\begin{figure}[H]
    \centering
    \includegraphics[width=0.5\linewidth]{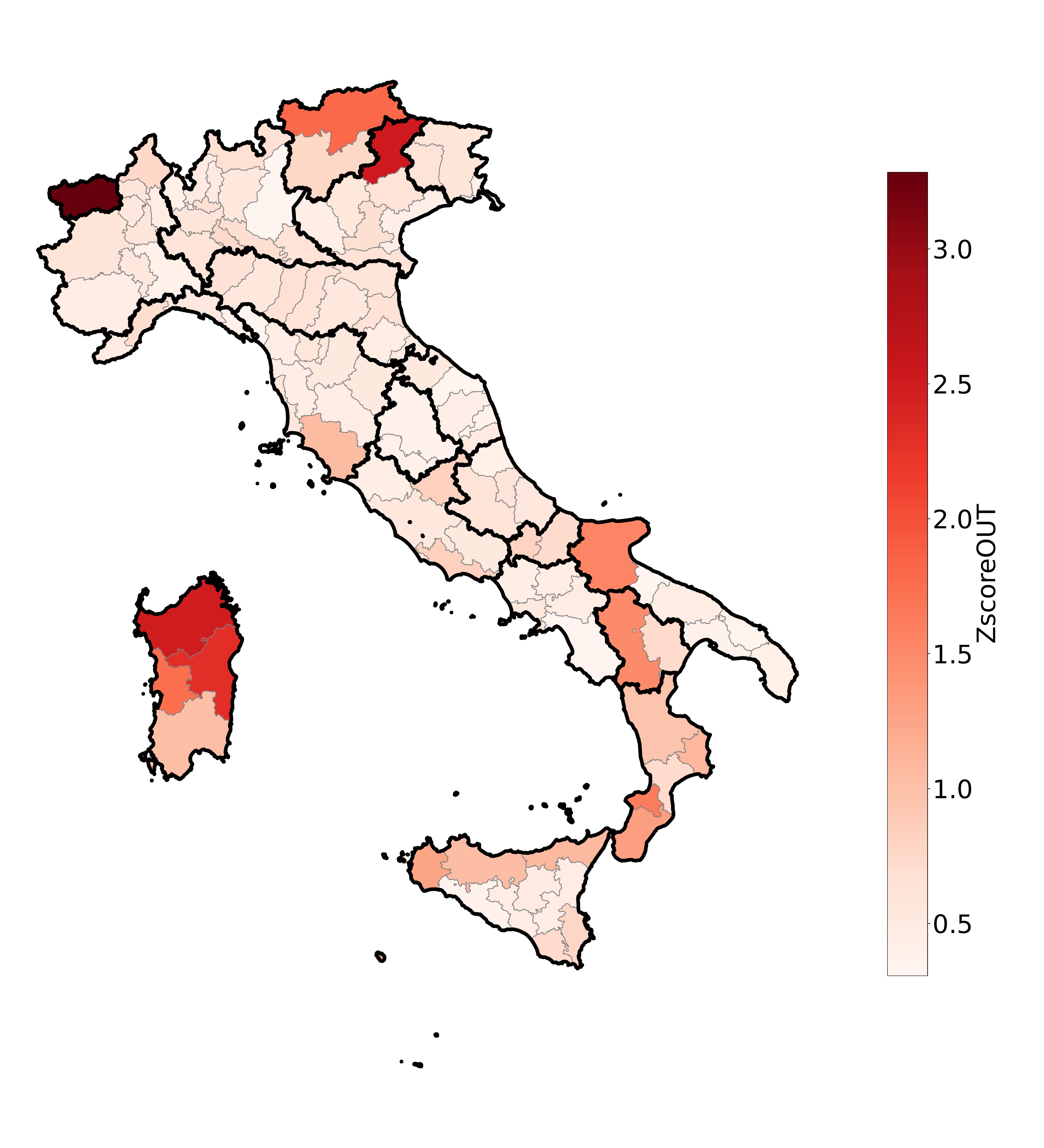}\includegraphics[width=0.5\linewidth]{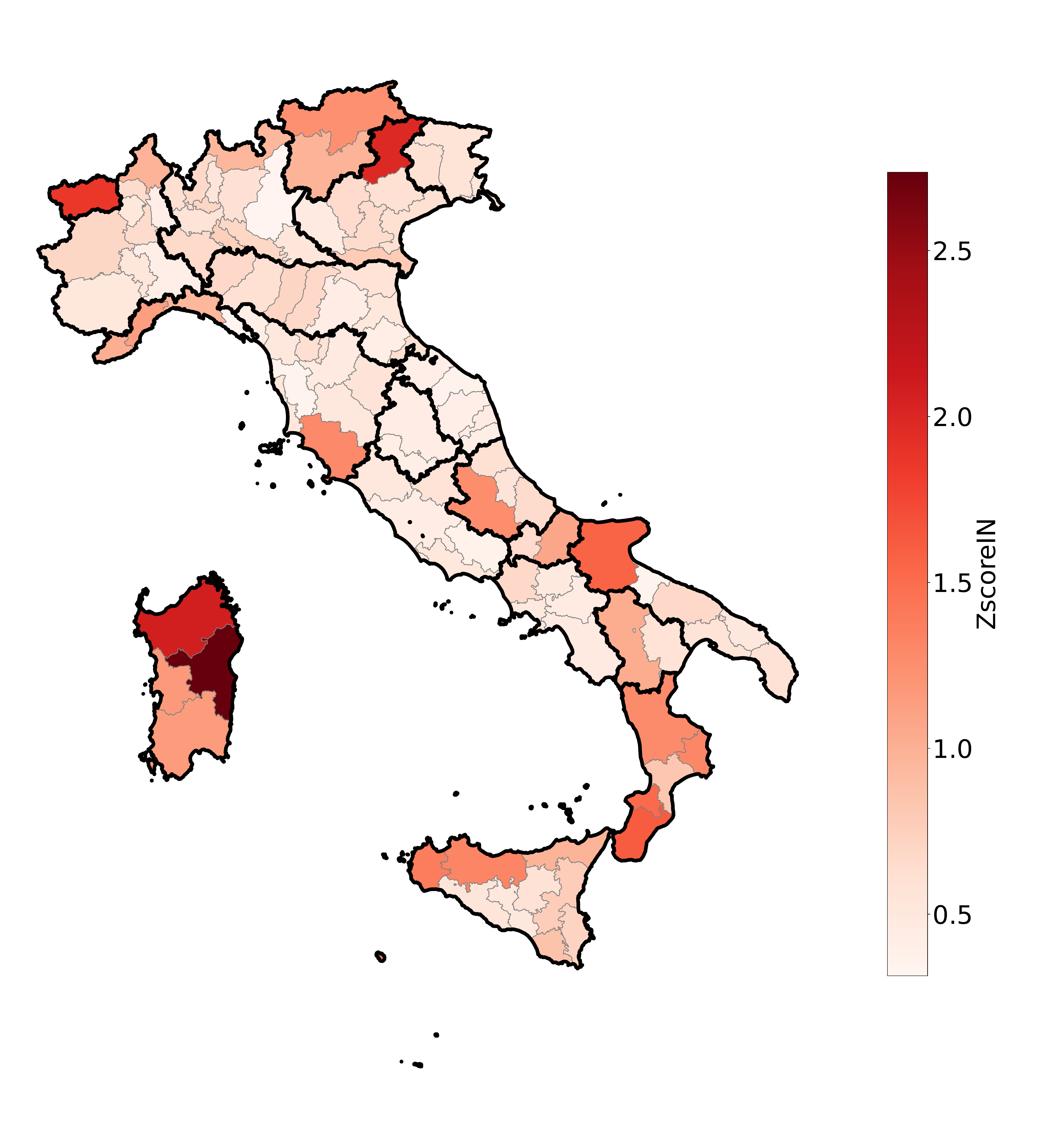}\\
    \caption{Map of the 2-year average Z score by province. The Z scores defined Eq.~\ref{eq:Zscore} are the average fluctuations by provinces of the probability of moving in (Z score IN) and out (Z score OUT) of the province.}
    \label{fig:Zscore}
\end{figure}

In Fig.~\ref{fig:Zscore}, we show at the top a map with the $Z$ score for the outgoing probabilities, and on the bottom one for ingoing probabilities for each province.

\section{In and outgoing probability for each spatial cluster}
\begin{figure}[H]
    \centering
    \includegraphics[height=0.9\textheight]{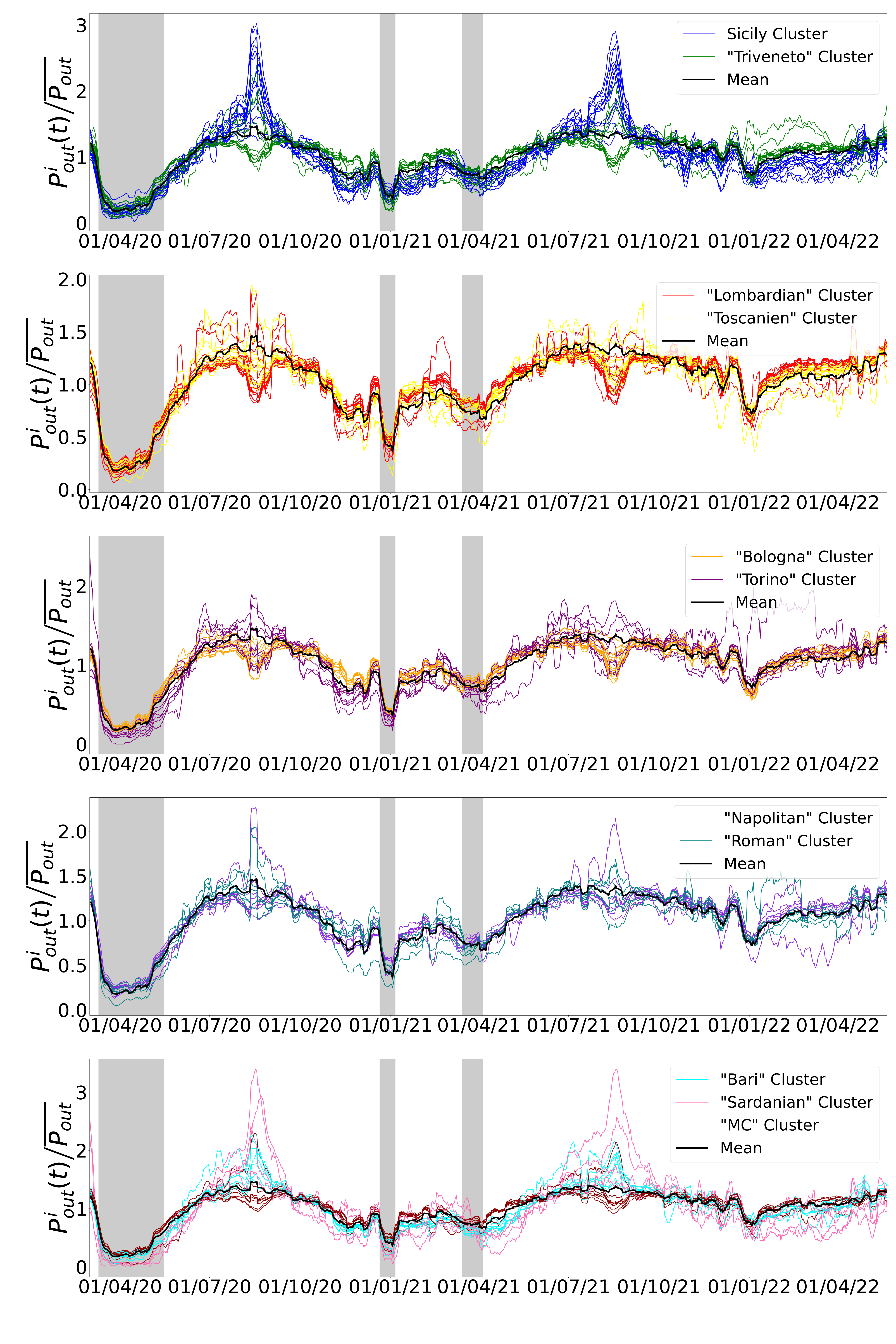}
    \caption{Outgoing probability for each cluster, the colors correspond to the clusters of the non-confined case using GMC.}
    \label{fig:OUT}
\end{figure} 

\begin{figure}[H]
   \centering
   \includegraphics[height=0.9\textheight ]{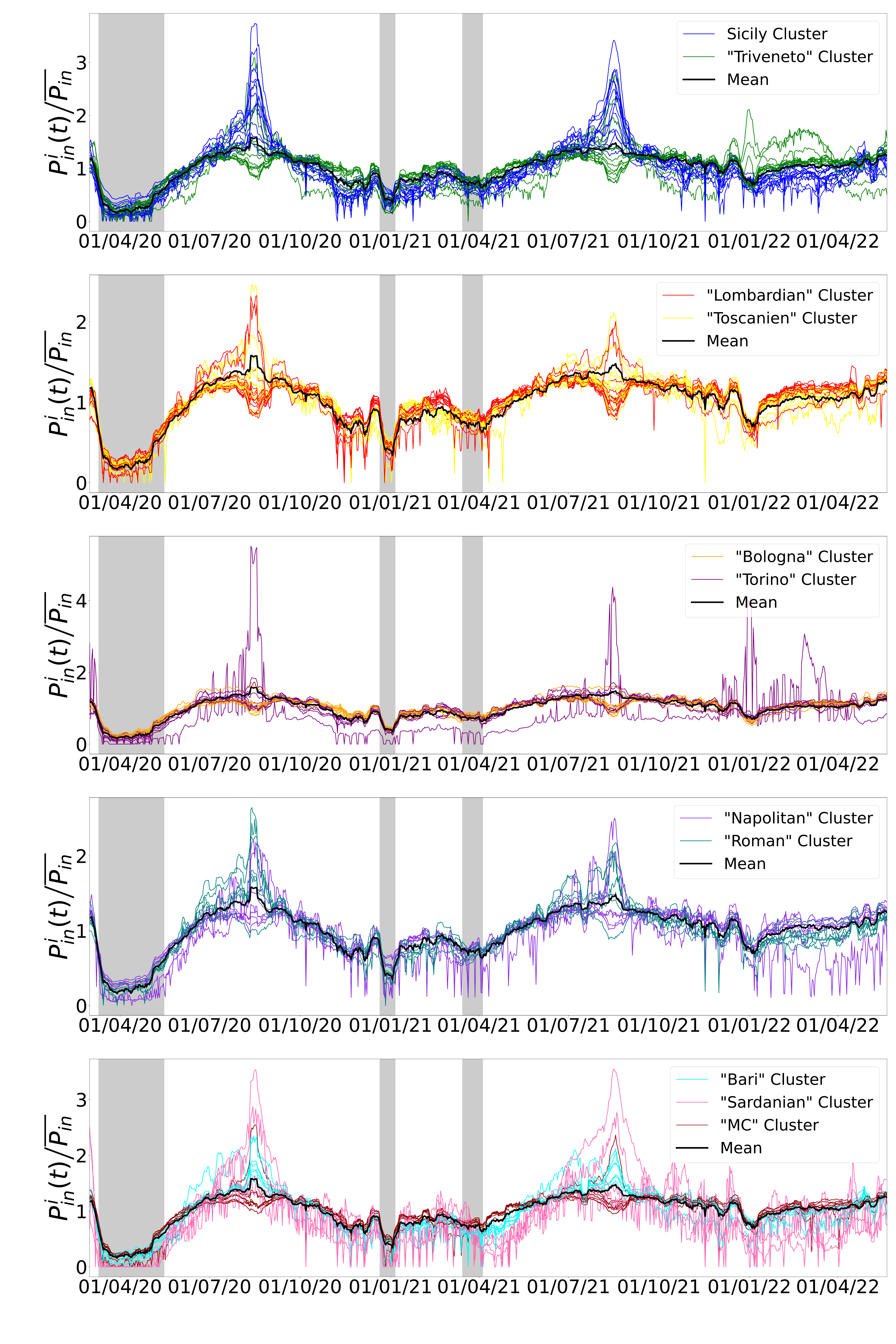}
   \caption{Ingoing probability for each cluster the colors correspond to the clusters of the non-confined case using GMC .}
   \label{fig:IN}
\end{figure}

\section{Network representation of the spatial partition with the two clustering methods}
We display here the full network visualization of the two most representatives obtained for the two temporal clusters, the widths of the links are proportional to the logarithm of the transition probability. 
\begin{figure}[H]
    \centering
    \title{Greedy Modularity Communities (GMC)}
    \includegraphics[width=\linewidth ]{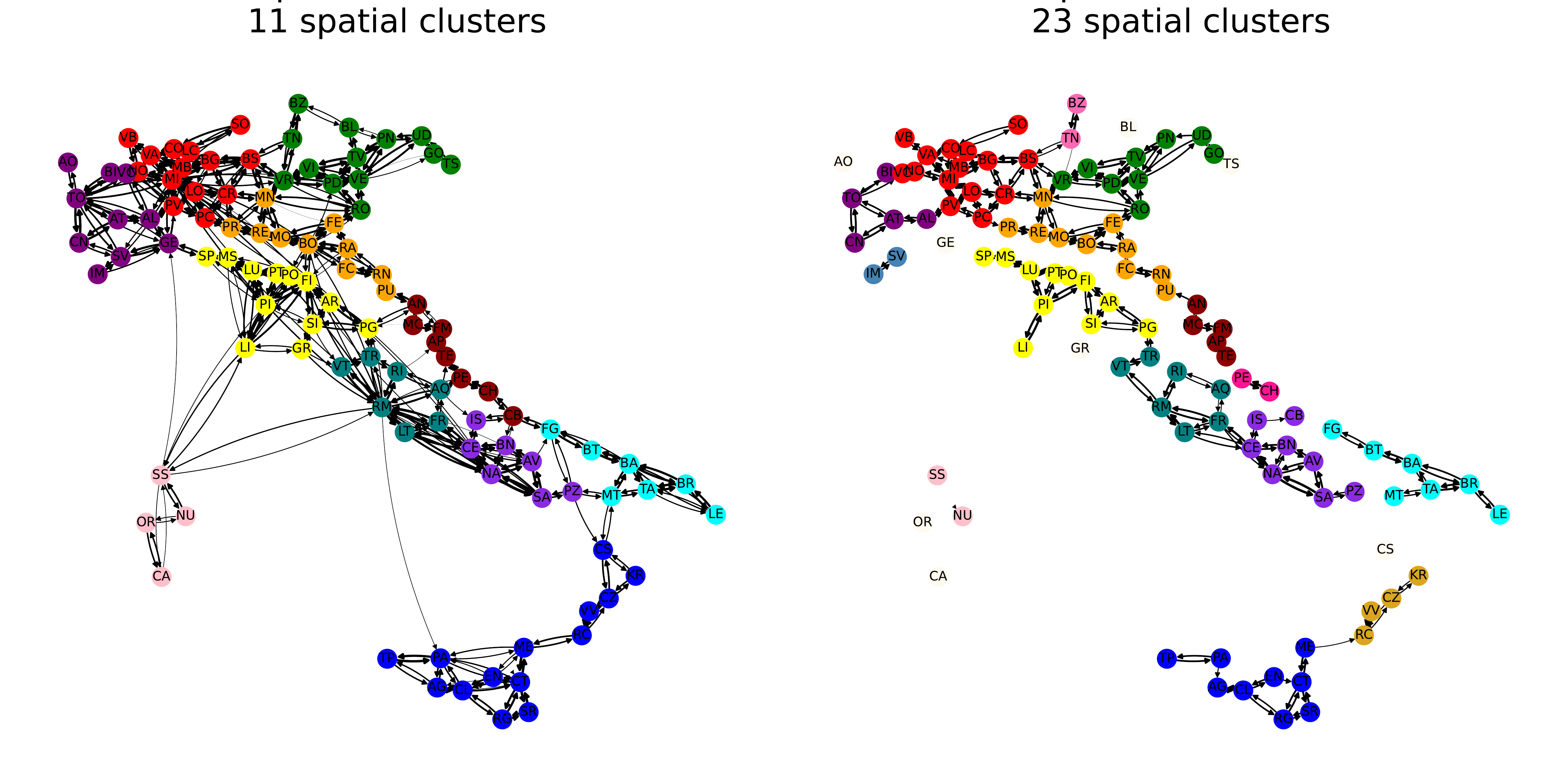 }
    \caption{Directed graph representation of the most representative matrices for non-confined cluster $C_0$ (top) and confined one $C_1$ (bottom) the optimal clustering in using the greedy modularity method.  }
    \label{fig:GraphGreedy}
\end{figure} 
\begin{figure}[H]
    \centering
    \title{Critical Variable Selection (CVS)}
    \includegraphics[width=0.9\linewidth ]{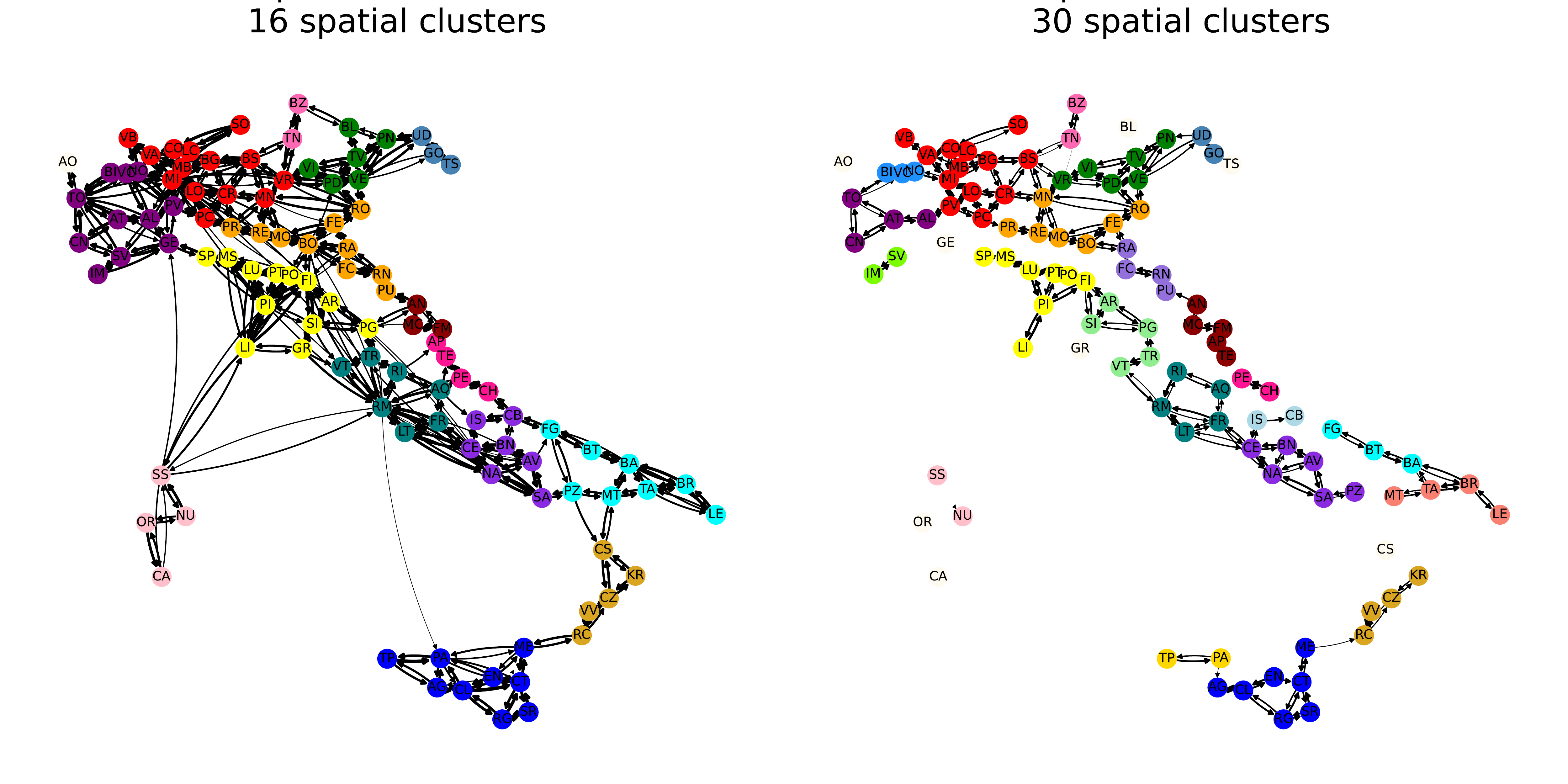 }
    \caption{Directed graph representation of the most representative matrices for non-confined cluster $C_0$ (top) and confined one $C_1$ (bottom) the optimal clustering in using the critical variable selection method. }
    \label{fig:GraphResRel}
\end{figure}

\end{document}